\documentclass[%
preprint,
 amsmath,amssymb,
 aps,
]{revtex4-1}
\usepackage{graphicx}%
\usepackage{multirow}%
\usepackage{amsfonts}%
\usepackage{amsthm}%
\usepackage{mathrsfs}%
\usepackage[title]{appendix}%
\usepackage{xcolor}%
\usepackage{textcomp}%
\usepackage{booktabs}%
\usepackage{listings}%
\usepackage{color}
\usepackage{epsfig}
\usepackage{latexsym}
\usepackage{float}
\usepackage{upgreek}
\usepackage{geometry}
\begin{document}
%\begin{titlepage}
\title{Memory of shear flow in soft jammed materials}
\author{H. A. Vinutha$^{1}$$^{\ast}$}
\author{Manon Marchand$^{2}$}
\author{Marco Caggioni$^{3}$}
\author{Vishwas V. Vasisht$^{4}$}
\author{Emanuela Del Gado$^{1}$}
\author{Veronique Trappe$^{2}$$^{\dag}$}
 \affiliation{1. Department of Physics, Institute for Soft Matter Synthesis and Metrology, Georgetown University, Washington, DC, USA}
 \email{vh163@georgetown.edu}
 \affiliation{2. Department of Physics, University of Fribourg, Fribourg, Switzerland}
 \email{veronique.trappe@unifr.ch}
 \affiliation{3. Complex Fluid Microstructures, Corporate Engineering, Procter \& Gamble Company, West Chester, Ohio 45069}
 \affiliation{4. Department of Physics, Indian Institute of Technology Palakkad,
Nila Campus, Kanjikode, Palakkad 678623, Kerala, India}

% word limit 600; currently 292
\begin{abstract}
{\bf Cessation of flow in simple yield stress fluids results in a complex stress relaxation process that depends on the preceding flow conditions and leads to finite residual stresses. To assess the
microscopic origin of this phenomenon, we combine experiments with largescale computer
simulations, exploring the behavior of jammed suspensions of soft repulsive particles. A spatio-temporal analysis of microscopic particle motion and local particle configurations reveals two contributions to stress relaxation. One is due to flow induced accumulation of elastic stresses in domains of a given size, which effectively sets the
unbalanced stress configurations that trigger correlated dynamics upon flow cessation. This scenario is supported by the observation that the range of spatial correlations of quasi-ballistic displacements obtained upon flow cessation almost exactly mirrors those obtained during flow. The second contribution
results from the particle packing that reorganize to minimize the resistance to flow by decreasing the number of locally
stiffer configurations. 
Regaining rigidity upon flow cessation then effectively sets the
magnitude of the residual stress. Our findings highlight that flow in yield stress fluids can be seen as a training process during which the material stores information of the flowing state through the development of domains of correlated particle displacements and the reorganization of particle packings optimized to sustain the flow. This encoded memory can then be retrieved in flow cessation experiments.}

%Cessation of flow in simple yield stress fluids results in a complex stress relaxation process that depends on the preceding flow conditions and leads to residual stresses.  To assess the microscopic origin of this phenomenon, we combine experiments with largescale computer simulations of jammed suspensions of soft repulsive particles.  A spatio-temporal analysis of microscopic particle motion and local particle configurations reveals two contributions to stress relaxation. One is due to flow induced accumulation of elastic stresses in domains, which effectively sets the unbalanced stress configurations that trigger correlated dynamics upon flow cessation. This scenario is supported by the observation that the range of correlations of fast ballistic displacements obtained upon flow cessation almost exactly mirrors those obtained during flow. The second contribution results from the particle packing that reorganize to minimize the resistance to flow by decreasing the number of locally rigid configurations. Regaining rigidity upon flow cessation then effectively sets the magnitude of the residual stress.  Our findings highlight that flow in yield stress fluids can be seen as a training process during which the material stores information of the flowing state through the development of domains of correlated particle displacements and the reorganization of particle packings optimized to sustain the flow. This encoded memory can then be read out in flow cessation experiments.  }

\end{abstract}
%\keywords{keyword1, Keyword2, Keyword3, Keyword4}

\maketitle

\section{Introduction}\label{sec1}
Soft jammed materials are suspensions of soft, deformable particles that are packed above the jamming transition. At these concentrations, the particles are in contact, interacting elastically with each other, and the material classifies as a weak solid. Despite its intrinsic solid properties, a soft jammed material can be continuously sheared without fracturing, which generates a stress ($\sigma$) that is
above a threshold stress, known as the dynamic yield stress ($\sigma_y$). Indeed, soft jammed materials belong to a broader class of materials called yield stress fluids \cite{nguyen1992measuring,bonn2017yield}, which encompasses foams, creams,
cement paste, etc. The use of these systems generally relies on their ability to flow under the application of either, a large enough stress or a constant shear rate \cite{nelson2019designing,nelson2020embedded}. This is why they can be conveniently pumped through pipes, squeezed out of tubes, spread on surfaces, and/or molded into a given shape.

To explore the consequences of flow history on the properties of yield stress fluids, experiments and simulations have been devised to measure stress relaxation upon flow cessation \cite{mohan2015build,hendricks2019nonmonotonic,sudreau2022residual,vasisht2022residual,sudreau2022shear}. This test
consists of first driving the system to a steady flow, where the stress remains constant in time. The strain rate is then set to zero, the strain maintained constant, and stress relaxation  measured as a function of time. Both the time scales of stress relaxation as well as the final residual stress reached at the
end of the relaxation process have been found to depend on the shear parameters setting the flow
conditions prior to flow cessation. However, an understanding of the processes leading to this relation remained elusive.

In this paper we explore the microscopic origin of this phenomenon combining experiments and simulations using highly packed systems of soft particles suspended in a continuous medium of varying viscosity. Our investigations reveal that flow encodes a memory of two processes. The first one relates to the system's response to the continuously increasing strain by forming finite-sized domains. Within these domains elastic energy is stored, evidenced by short-time non-affine particle displacements that are both directed and highly correlated within the domains. This process depends on shear rate and is independent of the viscosity of the suspending medium. The elastic load is isotropic, but locally unbalanced, leading upon flow cessation to fast quasi-ballistic displacements that exhibit spatial correlations mirroring those observed under flow. The second process relates
to a reconfiguration of the particle packing under flow, which leads to locally less stiff 
configurations, the overall stiffness being a function of the viscous stress experienced during flow. Upon flow cessation the particle packing evolves during stress relaxation until it reaches a rigidity level that satisfies the
conditions of mechanical stability; the time scale to reach this condition then sets the magnitude of residual stress. Our findings effectively disclose the origin of flow memory observed upon flow cessation and expose that flow cessation tests are efficient means to gain insight into the flow behavior of yield stress fluids \cite{nicolas2018deformation,vasisht2018rate,khabaz2020particle,aime2023unified,bandyopadhyay2010stress,barik2022origin,ballauff2013residual,bhattacharyya2023nature}.

\section{Results and Discussion}

\begin{figure}[htbp!]
%\centering
\includegraphics[scale=0.9]{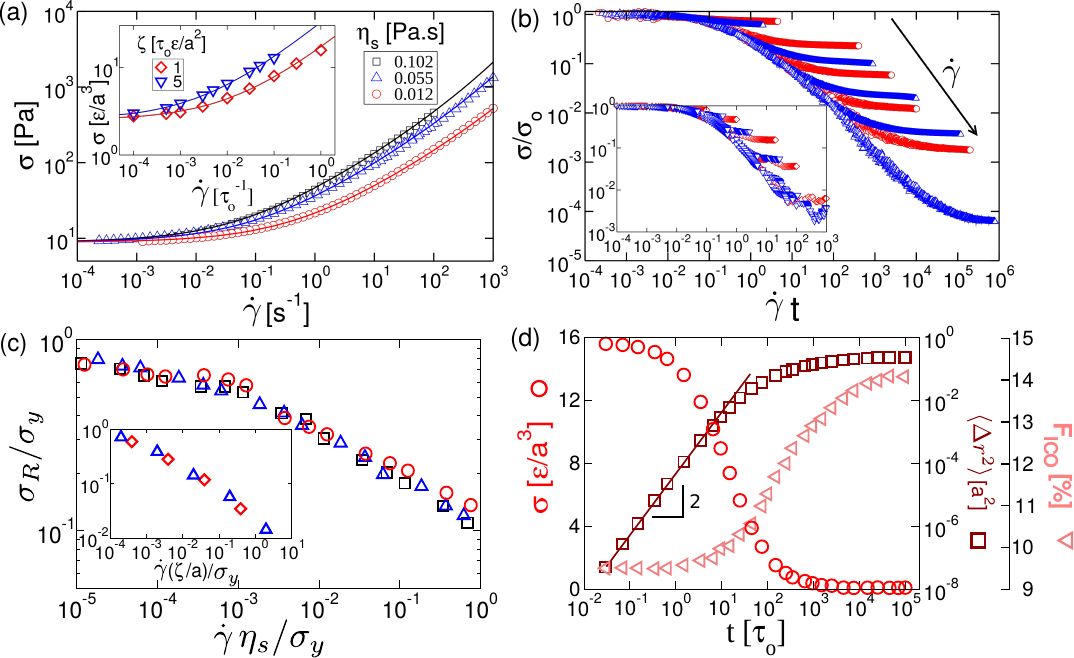}
\caption{\label{Fig1_expt} {\bf Evidence of two distinct contributions governing stress relaxation upon flow cessation.
} {\bf(a)}  Shear rate dependent stress (flow curves).  Main graph: experimental results obtained for different solvent viscosities.
%$T = 10^{\circ}$C/ $\eta_s = 0.102$  Pa s (black squares), $T = 20^{\circ}$C/ $\eta_s = 0.055$  Pa s (blue triangles up) and $T = 50^{\circ}$C / $\eta_s = 0.012$ Pa s  (red circles). 
Lines through the data are fits to the three component model \cite{caggioni2020variations}; the resulting yield stress is $\sigma_y=8.9$ Pa. Inset: results obtained in simulations for two damping coefficients $\zeta$.
%$\uptau_o \epsilon/a^2$ (blue triangles down) and $\zeta = 1$ $\uptau_o \epsilon/a^2$ (red diamonds).
Lines through the data are fits to the Herschel-Bulkley model \cite{bonn2017yield} fixing the yield stress value to $\sigma_y=2.48 \epsilon/a^3$, as obtained from quasi-static shear simulations. {\bf(b)} Stress relaxation upon flow cessation. The stress is normalized by the initial value $\sigma_o$ and the time is normalized with the inverse of the shear rate $\dot{\gamma}$. Main graph: selected experimental results obtained for $\eta_s = 0.055$  Pa s (blue triangles up) and $\eta_s = 0.012$ Pa s  (red circles).  From top to bottom the $\dot{\gamma} = 10^{-3}, 10^{-1}, 10^{0}, 10^{1}, 10^{2}$ s$^{-1}$. Inset: results obtained in simulations for $\zeta = 5$ $\uptau_o \epsilon/a^2$  (blue triangles down) and $\zeta = 1$ $\uptau_o \epsilon/a^2$  (red diamonds). From top to bottom $\dot{\gamma} = 10^{-4}, 10^{-3}, 10^{-2}, 10^{-1}, 10^{0}$ $\uptau_o^{-1}$. {\bf(c)} Residual stress normalized by the yield stress as a function of the viscous stress experienced during shear.  Main graph: experimental results obtained for the three conditions described in (a). Inset: results obtained in simulations for the two conditions described in the inset of (a).  {\bf (d)} 
Stress relaxation upon flow cessation for $\dot{\gamma} = 1.0$ $\uptau_o^{-1}$ and $\zeta = 1$ $\uptau_o \epsilon/a^2$ compared to the corresponding time dependences of the mean squared displacement $\langle\Delta r^2\rangle$ and the fraction of Voronoi cells with icosahedral configuration $F_{ICO}$.}
\end{figure}

%---Figure 1 ----
As introduced above, the aim of this work is to address how the application of a continuous shear rate imprints memory into a soft jammed material. To gain an understanding of the effect of the shear rate applied independently of the resulting stress, we explore the flow behavior of a densely packed microgel system composed of Carbopol dispersed in propylene glycol and vary the viscosity of the dispersing medium by varying the temperature. The details of the sample preparation and sample characteristics are given in the Methods section. 
As shown in the main graph of Fig. \ref{Fig1_expt}(a), our system exhibits the typical flow characteristics of a simple yield stress fluid. While the stress increases monotonically in the range of high shear rates, the stress is almost shear rate independent at low shear rates. Clearly, the system can only sustain a steady flow when the stress exceeds the dynamic yield stress $\sigma_y$, defined as the value of stress in the limit of $\dot{\gamma} \rightarrow 0$. The yield stress is independent of the solvent viscosity, which denotes that $\sigma_y$ is exclusively set by the intrinsically solid characteristics of the dense ensemble of soft spheres. By contrast, at higher rates, the stress sensitively depends on the solvent viscosity $\eta_s$. For a given shear rate the stress increases with increasing $\eta_s$, which reflects that the stress generated at finite shear rates also depends on viscous dissipation in the continuous phase.

To gain insight in the shear rate dependent states, we perform flow cessation experiments at various shear rates along the flow curve. These tests consist of stopping the shear after reaching steady flow, and to subsequently measure stress relaxations while holding the strain constant. Consistent with previous work \cite{mohan2015build,ballauff2013residual,vasisht2022residual}, we find that the stress relaxation process upon flow cessation sensitively depends on the shear rate used to prepare the system. The higher the preshear rate the faster the stress relaxation and the lower the stress reached in the long-time limit of the stress relaxation test (see SM Fig. 1). 
Normalizing the stress by its initial value $\sigma_o$ at $t=0$s, and the time by the preshear rate, highlights an initial decay common to all flow conditions, independent of solvent viscosity, as shown in Fig. \ref{Fig1_expt}(b).  
Such behavior indicates that the initial fast stress relaxation is not strictly related to a structural relaxation process that would require a significant relative displacement between neighboring particles, which should also depend on solvent viscosity. %It suggests that the short-time behavior relates to an {\it elastic} relaxation, involving cooperative particle displacements.

At longer times, however, stress relaxation deviates from the unique behavior set by the preshear rate, the stress becoming eventually time-independent. This final stress plateau is commonly referred to as residual stress \cite{ballauff2013residual,mohan2015build,vasisht2022residual} and reveals that elastic loading is never suppressed during shear. Indeed, a system subjected to a steady shear is forced to continuously reconfigure. However, in yield stress fluids, reconfiguration occurs intermittently, which allows for constant elastic reloading \cite{khabaz2020particle,vasisht2018rate,song2022microscopic,hwang2016understanding}. Naturally, as the strain is held constant during the flow cessation test, a net elastic load remains after the initially imbalanced local stresses relaxed below the local yield stresses. In contrast to the short-time relaxation, the residual stress is not solely set by the preshear rate. As denoted in Fig. \ref{Fig1_expt}(b), preparing a system with a given shear rate will result in different residual stresses upon varying the solvent viscosity: the lower the solvent viscosity the larger the residual stress. The residual stress ${\sigma}_R$ depends in fact on the viscous stress $\dot{\gamma} {\eta}_s$; indeed, reporting ${\sigma}_R$ obtained at different viscosities as a function of $\dot{\gamma} {\eta}_s$ results in a unique master curve,  as shown in Fig. \ref{Fig1_expt}(c)  (for the unscaled data, see SM Fig. S1). Taking the residual stress as the relevant characteristic of the final stage of the stress relaxation process, the dependence on solvent viscosity indicates that the final stage involves a structural relaxation that requires the particles to reorganize with respect to each other and thus depends on viscous dissipation. 

%\cite{ballauff2013residual,song2022microscopic,sudreau2022residual,lidon2017power,dallari2020microscopic,bhattacharyya2023nature,pamvouxoglou2021stress,ballauff2013residual,mohan2013microscopic,mohan2015build,ness2016two,vasisht2022residual,barik2022origin}.  

These findings clearly indicate that stress-relaxation upon flow cessation depends on the shear parameters setting the flow conditions prior to flow cessation. To understand the microscopic origin of this dependence, we perform molecular dynamics simulations for a jammed packing of non-Brownian soft repulsive particles dispersed in an implicit solvent, in which they experience a drag force dependent on the solvent viscosity \cite{vasisht2020computational,vasisht2020emergence}. More detailed information on the numerical model and simulations is provided in the Methods section. The characteristics of the flow curves and the flow cessation results obtained in the simulations with different drag coefficients $\zeta$ qualitatively agree with the experimental findings obtained at different $\eta_s$, as shown in the insets of Figs.~\ref{Fig1_expt}(a), (b), and (c). In particular, the scaling behaviors denoted in the insets of Figs.~\ref{Fig1_expt}(b) and (c) confirm that the short-time and long-time characteristics of stress relaxation are predominantly governed by respectively the shear rate and the viscous stress experienced during shear.  

To gain a first microscopic understanding of these two distinct relaxation regimes we characterize the particle dynamics during stress relaxation by determining the mean squared displacement $\langle \Delta r^2 \rangle$, and we assess the structural evolution by decomposing the evolving particle packings into Voronoi cells, which enables us to determine the fraction of Voronoi cells that have an icosahedral shape, $F_{ICO}$. As shown in previous work \cite{pinney2016structure,vasisht2020emergence,vasisht2020computational}, icosahedrally shaped Voronoi cells identify locally stiffer regions in particle packings. The specific relevance of this particular structural parameter will be discussed later; for the current discussion, let us note that other structural parameters, like the mean number of particle contacts \cite{mohan2015build,mohan2013microscopic,cuny2021microscopic}, have been shown to display similar hallmarks in their temporal evolution as those observed in the temporal evolution of $F_{ICO}$.

As an example, we show the time dependence of $F_{ICO}$  obtained at a high shear rate $ \dot{\gamma} = 1$ $\uptau_o^{-1}$ and $\zeta = 1$ $\uptau_o \epsilon/a^2$, in comparison to the time dependence of respectively the mean squared displacement and the stress in Fig. \ref{Fig1_expt}(d). Upon flow cessation, $F_{ICO}$  remains essentially unchanged within a short time interval, while the stress drops by about $30\%$. During this initial period the mean squared displacement increases with the square of time. This highlights that the initial relaxation, which solely depends on shear rate, independent of viscosity, is associated to quasi-ballistic motion without major changes in particle configuration. By contrast, a significant reorganization of the particle packing occurs in the final stage of stress relaxation.    

\begin{figure}[htbp!]
%\centering
\includegraphics[scale=0.6]{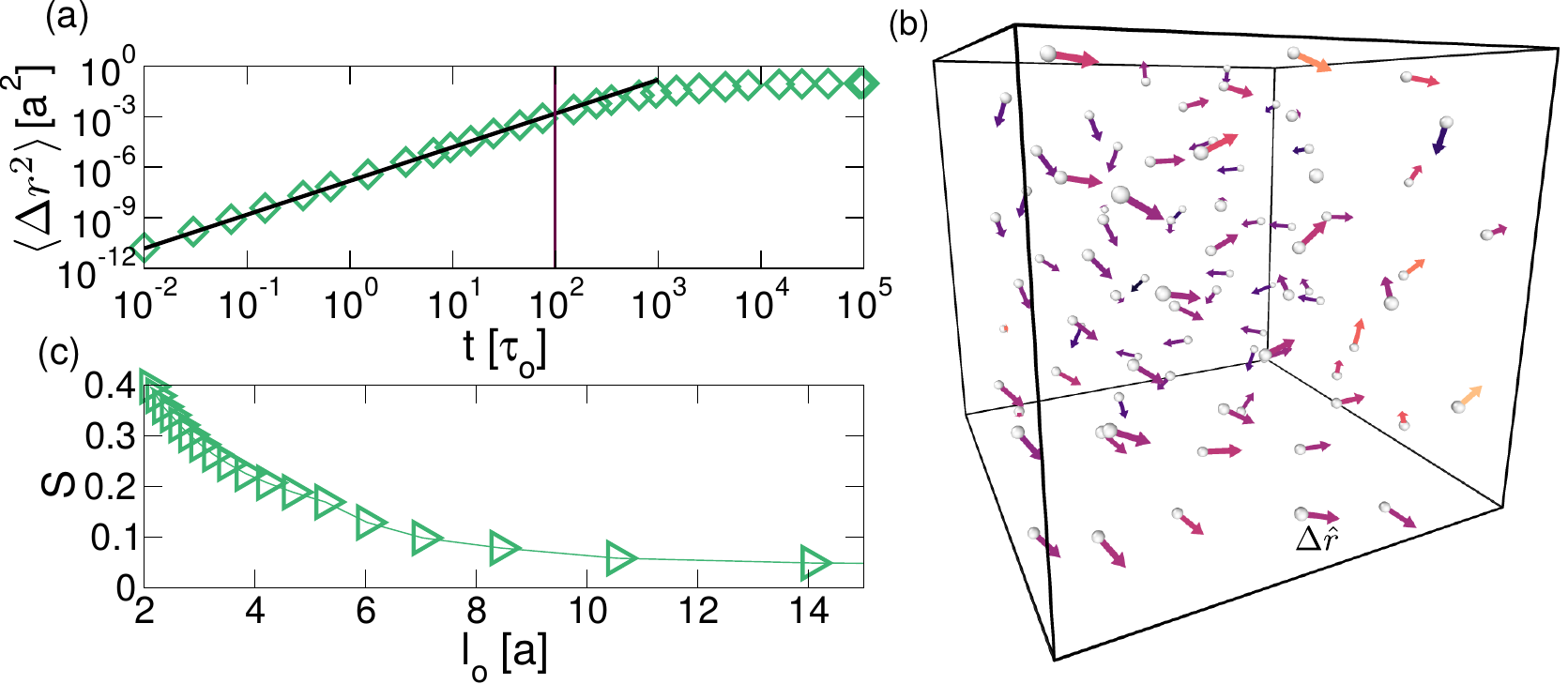}
\caption{\label{ball} {\bf Evidence of isotropically oriented  domains of correlated dynamics.} Example of spatial analysis of particle displacements upon flow cessation ( $\dot{\gamma} = 10^{-2}$ $\uptau_o^{-1}$ and $\zeta = 1$ $\uptau_o \epsilon/a^2$). {\bf(a)} Time dependence of mean squared displacement, the vertical line denoting the delay time used for the spatial analysis of the particle displacements. {\bf (b)} ($4a\times 4a \times 4a$) sub-volume of the simulation box showing the position of the particles center of mass upon flow cessation as a dot, the arrows representing the unit vectors of the subsequent particle displacements  $\Delta \hat{r}$. The arrows are colored according to the unit vector direction. 
{\bf(c)} Scalar order parameter $S$ as a function of the size of the sub-volume of the simulation box $l_o$. }
\end{figure}

In absence of inertial effects (see Methods), the ballistic displacements can be ascribed to a partial release of particle contact deformations that were set during flow \cite{mohan2013microscopic,khabaz2020particle,mohan2015build}. 
This quasi-elastic relaxation should be largely independent of the solvent viscosity, which is consistent with the initial scaling of the stress relaxation process discussed above (Fig.~\ref{Fig1_expt}(b)). 
%----Figure 2----
More convincing evidence for the elastic origin of the initial stress relaxation is found, by analysing the spatial configuration of the particle displacements, where we find 
that the particle displacements are not only directional, but also highly correlated. To illustrate this, we show as an example the spatial configuration of the displacement unit vectors observed within a sub-volume of the simulation box obtained for $ \dot{\gamma} = 10^{-2}$ $\uptau_o^{-1}$ and $\zeta = 1 \uptau_o \epsilon/a^2$ in Figure \ref{ball}(b); the time delay is here chosen to correspond to the time window over which we observe $\langle \Delta r^2 \rangle \propto t^2 $, marked by a vertical line in Figure \ref{ball}(a), and the sub-volume chosen has a linear size of $\approx 4$ particle diameters. To highlight the degree of correlation we assign a given color to the displacement vectors with the same pointing direction. Domains in which particles move along the same direction are clearly identified, supporting the idea that the process governing the initial stress relaxation relates to an elastic relaxation of contacting particles that have been compressed during shear flow. Performing this analysis throughout the simulation box reveals that many such domains coexist, the direction of decompression changing from one domain to another. As a consequence, an analysis of the particle displacements along different directions yields that the overall particle displacements is isotropic, as shown in SM Fig. S2 and observed in previous investigations \cite{mohan2015build,mohan2013microscopic,khabaz2020particle}. 

To properly quantify the degree of alignment of the particle displacements, we determine a scalar order parameter $S$ defined as the largest positive eigenvalue of the tensor obtained from the displacement directors, which is similar to the approach used to describe the degree of orientation in nematic liquid crystals  \cite{chaikin1995principles}. 
The value of $S$ is zero when the unit vectors are randomly oriented and $S=1$ when all unit vectors point in the same direction. This analysis is performed varying the linear size $l_o$ of sub-volumes of the simulation box. As shown in Fig.~\ref{ball}(c), $S$ decreases with increasing $l_o$ and saturates to $0.05$ beyond $l_o=12$. This fully supports our previous assessment, namely that the system contains correlated domains of finite size, the pointing direction of the displacement vectors differing from one domain to another.

\begin{figure*}[h!]
%\centering
\includegraphics[scale=0.73]{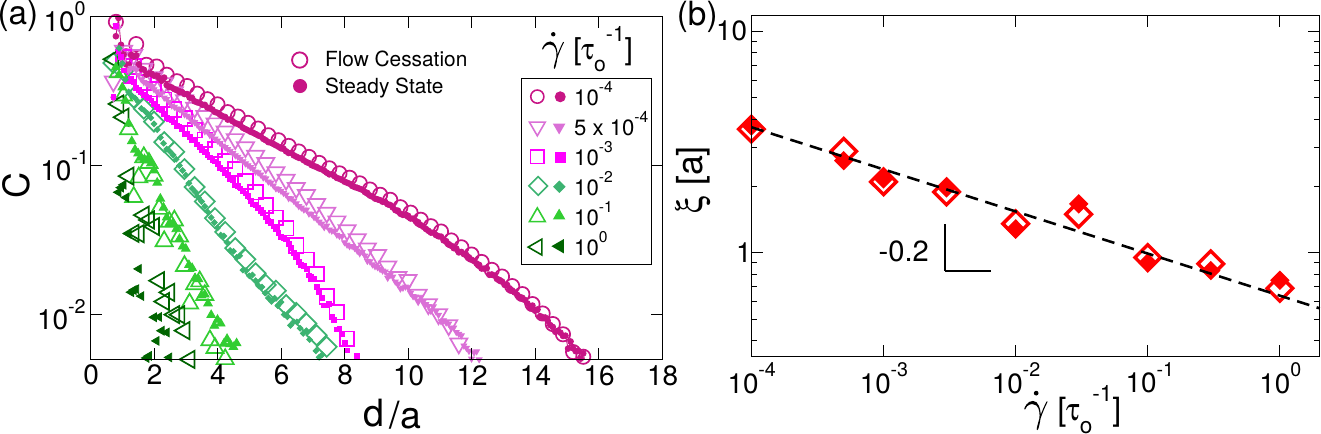}
\caption{\label{corel} {\bf Memory of spatial correlations of particle displacements.} {\bf(a)} Selected examples of the spatial correlation function of dynamical fluctuations $C$ obtained for $\zeta = 1$ $\uptau_o \epsilon/a^2$. The closed symbols denote data obtained just before flow cessation and the open symbols denote the data obtained just after flow cessation.  From top to bottom $\dot{\gamma} = 10^{-4}, 5 \times 10^{-4}, 10^{-3}, 10^{-2}, 10^{-1}, 10^{0}$ $\uptau_o^{-1}$. {\bf(b)} Shear rate dependence of the correlation length determined from the decay rate of $C$. The dashed line is a power-law fit to the data with exponent $-0.2$.}
\end{figure*}

%----Figure 3----
The existence of dynamically correlated domains, observed upon flow cessation, raises the question of their origin and their dependence on shear rate. To address this question, we investigate the correlation in particle displacements at different shear rates. As we expect not only the displacement direction to be correlated but also the displacement magnitude, we determine for each particle $i$ the magnitude of the displacement $\Delta r_i$ observed within the time interval corresponding to that at which $\langle \Delta r^2 \rangle \propto t^2 $, and we assess its fluctuation over the average displacement, $\delta u_i = \Delta r_i - \langle \Delta r \rangle$. As a measure for correlation we then compute the spatial correlation function $C(d) = \frac{\langle \delta u_i \delta u_j \rangle}{\langle \delta u^2 \rangle}$ for all pairs of particles $i$ and $j$ separated by a distance $d$. This spatial correlation function strongly depends on the shear rate applied prior to flow cessation. Particles moving much more or much less than the average are correlated across a given distance, this distance being a decreasing function of the preshear rate, as shown in Fig. \ref{corel}(a) (open symbols). 
Strikingly, an equivalent analysis of the spatial correlation of non-affine particle displacements observed during shear flow
reveals a direct correspondence between the dynamical correlations induced during shear flow and those observed upon flow cessation; as shown in Fig. \ref{corel}(a), $C(d)$ obtained for shear flow (closed symbols) almost perfectly superimpose $C(d)$ obtained for flow cessation (open symbols). Let us note, that for a given shear rate the delay time chosen to determine $\Delta r_i$ is the same for both experiments, shear flow and flow cessation. Indeed, this delay time actually covers the time window for which $\langle \Delta r^2 \rangle \propto t^2 $ in both experiments, a quasi-ballistic motion being also observed in the non-affine dynamics of the sheared system \cite{vasisht2018rate,khabaz2020particle}.   

The remarkable correspondence between dynamical correlations observed respectively during steady flow and upon flow cessation can be understood by considering that dynamics following flow cessation is a result of stress imbalances that are quenched in during flow. Within this framework, we postulate that stress fluctuations induced by correlated dynamics during flow are imprinted as stress imbalances upon flow cessation. These stress imbalances then serve as a source for dynamics during stress relaxation. Given that the spatial extent of stress fluctuations is set by the dynamically correlated domains during flow, the resulting stress imbalances driving correlated dynamics upon flow cessation reproduce the same correlation pattern. It is here worth noticing that 
flow cessation tests effectively capture the exact imprint of the sheared states just before flow cessation. Indeed, repetitions of a given flow cessation test in independent simulation runs show that $C(d)$ varies at large $d/a$ from one run to another, yet the excellent agreement between $C(d)$ obtained upon flow cessation and $C(d)$ obtained during flow is maintained, as long we compare the data collected just prior to and immediately after flow cessation (see SM Fig. S3). 
These observations underscore that subjecting jammed systems to a constant shear rate is akin to a training process, inherently encoding a memory of the sheared state through the presence of domains of correlated dynamics. Consequently, the flow cessation test can be used to gain information on the sheared state. This is of particular interest to experimental work aiming to assess correlations in the non-affine displacement under shear flow. Exploring this while applying a constant shear rate requires to account for affine displacements, which is a non trivial endeavor in experiments, while this information can be gained without any correction upon flow cessation.   
However, let us note that though the spatial correlations of non-affine displacements during flow and upon flow cessation are indistiguishable, the absolute magnitude of the displacements are not identical. The non-affine displacements are larger during flow than during stress relaxation upon flow cessation. Upon flow cessation particle displacements are thus distinct from those observed during shear flow, but the information of the spatial correlations of displacements between neighboring particles is preserved.  

As denoted in Fig. \ref{corel}(a), the initial decay of $C(d)$ can be approximated as an exponential. This enables us to determine a dynamical correlation length $\xi$, from the decay rate of $C(d)$ (see Methods). Theories of elastoplastic flow {\cite{picard2002simple,bocquet2009kinetic,nicolas2018deformation,ferrero2014relaxation,lin2018microscopic,benzi2021continuum} predict the emergence of dynamical correlations, which are expected to grow as the shear stress approaches the dynamical yield stress. Direct evidence of such correlations has been so far missing in both experiments and simulations, but are directly revealed in our analysis. Broadly consistent with theory {\cite{clemmer2021criticality,clemmer2021criticality1,lin2014scaling}, we find that the shear rate dependence of $\xi$ can be described by a power-law with an exponent $\simeq -0.2$, as shown in Fig. \ref{corel}(b). 
This provides compelling evidence that subjecting a jammed system to a steady shear flow requires the system to break down into finite domains to accommodate the continuously increasing deformation, thereby releasing elastic stresses through localized plastic events, while storing stresses within the domains.  Notably, the domain size increases with decreasing shear rate, indicating that a domain will essentially span the entire system at low enough shear rate \cite{nicolas2018deformation}.

Based on our discussion so far, one may be inclined to assume that stress relaxation upon flow cessation is entirely governed by dynamical heterogeneities encoding stress-imbalances during shear flow, these imbalances driving dynamics upon flow cessation until all local stresses dropped below the local yield stresses \cite{cuny2021microscopic,cuny2022dynamics}. However, shear flow does not only encode stress imbalances; it also alters the spatial particle configurations that need to adapt to the continuous flow. In particular, shear flow reduces the number of icosahedral configurations, i.e. the amount of locally stiffer regions; it thus effectively shifts the distribution of local yield stresses. Measuring local yield stresses directly from the microscopic configuration is highly challenging, the characterization of the particle packing through  $F_{ICO}$, however, provides semi-quantitative insight into this aspect of the problem.

As shown in Fig. \ref{Icosa}(a), the fraction of icosahedral Voronoi cells observed under steady shear flow $F_{ICO}^{S}$ systematically decreases with increasing shear rate.  Reminiscent of the scaling behavior observed for the residual stress, $F_{ICO}^{S}$ scales with the viscous stress (for the unscaled data see SM Fig. S4).  
The emerging picture is that the structural organization compatible with steady shear flow becomes increasingly incompatible with rigidity, i.e. mechanical stability at rest, as the viscous stress across the system increases. Consequently, the more the particle configuration has been altered with respect to that at rest, the more structural reorganization is needed to regain mechanical stability.

\begin{figure*}[h!]
\centering
\includegraphics[scale=0.75]{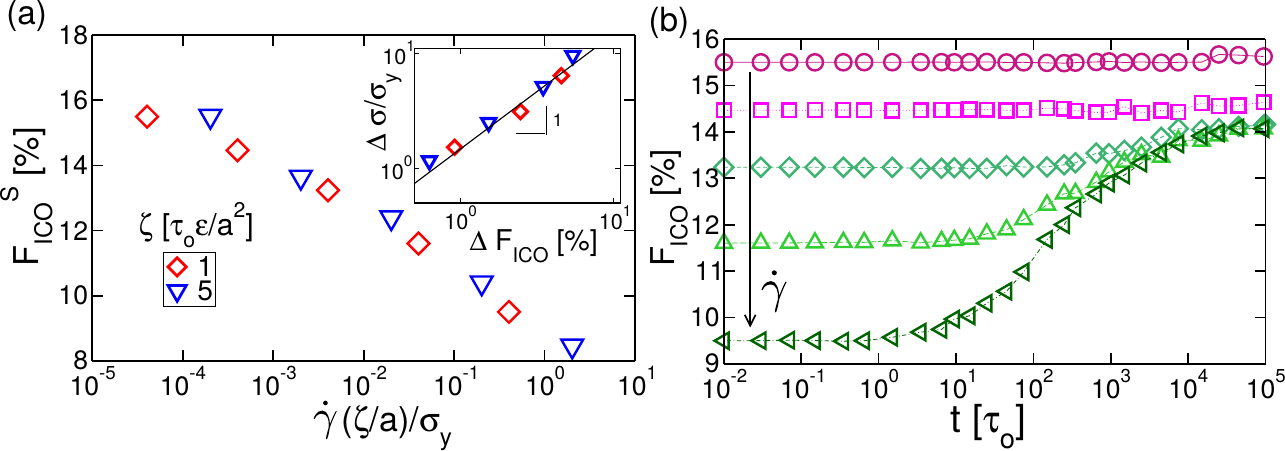}
\caption{\label{Icosa} {\bf Flow-induced weakening of particle configurations determining the magnitude of residual stresses upon flow cessation.} {\bf(a)} Main graph: fraction of Voronoi cells with icosahedral configuration obtained during steady shear $F_{ICO}^{S}$ as a function of the viscous stress normalized by the yield stress. Blue triangles down denote the data obtained for $\zeta = 5$  $\uptau_o \epsilon/a^2$, red diamonds denote the data obtained for $\zeta = 1$  $\uptau_o \epsilon/a^2$. Inset: magnitude of stress relaxation $\Delta \sigma$ normalized by the yield stress as a function of the magnitude of the changes in $F_{ICO}$ observed after application of larger shear rates (see Fig. \ref{Icosa}(b)).  
{\bf(b)}  Temporal evolution of $F_{ICO}$ upon flow cessation obtained for  $\zeta = 1$  $\uptau_o \epsilon/a^2$. From top to bottom: the preshear rate is $\dot{\gamma} = 10^{-4}, 10^{-3}, 10^{-2}, 10^{-1}, 10^{0}$ $\uptau_o^{-1}$. }
\end{figure*}

Indeed, for large enough viscous stresses, i.e. low enough $F_{ICO}^{S}$, $F_{ICO}$ increases during stress relaxation, as denoted in Fig. \ref{Icosa}(b). Clearly, for the stress relaxation to end after flow cessation, the system needs to regain a configuration compatible with a minimum rigidity, which appears to correspond to a minimum amount of icosahedrally packed domains. 
To assess this relation we determine the difference between the stress generated during flow and the residual stress $\Delta \sigma = \sigma_o - \sigma_R$, and the corresponding difference between $F_{ICO}^{S}$ and $F_{ICO}$ obtained at the end of stress relaxation $\Delta F_{ICO} = F_{ICO}^{S} - F_{ICO}^{final}$. As shown in the inset of Fig. \ref{Icosa}(a), both quantities relate linearly to each other, which strongly supports the idea that the magnitude of the residual stress is governed by conditions of mechanical stability that are well captured by the number of icosahedral configurations present in the system. Let us note that $F_{ICO}$ does not evolve during stress relaxation if the pre-shear rates are low, such that $F_{ICO}^S$ remains high enough. However, an inspection of the evolution of the structural configuration here reveals that, while $F_{ICO}$ does not increase, the positions of the icosahedrally packed regions change during stress relaxation, as shown for $\dot{\gamma} = 10^{-4} \uptau^{-1}$ in Movie 1 in the supplementary material. These findings suggest that the actual arrangement of icosahedrally packed regions in space may be another factor determining the mechanical stability of the system at rest, consequently affecting the magnitude of residual stress. These insights highlight outstanding questions to be the subject of future work.     

\section{Conclusion}

In conclusion, our work discloses evidence of flow memory in jammed packings of soft spheres. This memory leads to distinct features in the stress relaxation response upon flow cessation, denoting that flow cessation experiments are valuable means providing insight into the state of a system
subjected to a given shear rate. In the low shear rate limit, where the shear stress is of the order of the yield stress, memory is mainly encoded via dynamical correlations. Non-affine particle displacements are directed and highly correlated encoding long-range stress imbalances. These in turn trigger quasi-ballistic displacements upon flow cessation whose spatial correlations are indistinguishable to those observed under flow. As the shear rate is increased the range of correlated motion decreases, and the particle configurations evolve towards mechanically less stiff configurations. The reduced stiffness facilitates stress-relaxation and entails an evolution of the local particle configurations during stress relaxation towards a particle packing that is sufficiently rigid to warrant mechanical stability. The two contributions determining stress relaxation upon flow cessation, namely local stress imbalances and unstable particle configurations, are the determining parameter at different time scales. The relaxation of elastically loaded contact forces is the predominant process determining stress relaxation at short time and is set exclusively by the preshear rate. The rearrangements towards mechanically stable particle packings, in turn, determine the long-time relaxation towards a residual stress. 

Our work clearly exposes the microscopic origin of the relation between shear flow characteristics and stress relaxation upon flow cessation, and reveals how shear history is encoded into a soft jammed material.

\section{Methods}
\textbf{Experiments:} Our experimental system is a dispersion of $2 w \%$ Carbopol in propylene glycol. The Carbopol used is a  commercial polyacrylic acid microgel (974p (Lubrizol)), which is polydisperse in size, and has a mean radius on the order of $\approx10\mu m$. Based on the development of the shear modulus as a function of concentration (see SM Fig. S5), we estimate the volume fraction of our sample to be about 74\%. At this condition the microgels are in contact and deformed, but unlikely to be significantly compressed \cite{conley2017jamming,conley2019relationship}. 
The reason for choosing propylene glycol as dispersing medium is threefold. It is a fluid with a low vapor pressure (10.6~Pa at $T=20^{\circ}$C), such that evaporation is very slow compared to the duration of our experiments. It's viscosity ($\eta_s = 0.055~$Pa.s at $T=20^{\circ}$C) is high enough to allow the high shear rate range of a flow curve to fall within the experimentally accessible range of shear rates. Finally, it's viscosity varies by a factor of about 10 within a temperature range of $10-50^{\circ}$C ($\eta_s = 0.102-0.012$ Pa s), which enables us to assess the importance of the viscous stresses ($\eta_s \dot{\gamma}$) independently of the shear rate.

To prepare our system we mix $2 w \%$ of Carbopol (974p (Lubrizol)) into propylene glycol using a high shear mixer (Silverson LS M-A) operating for 5 min at $7000$rpm. The sample is then sonicated for $6$ hours in a sonicating bath thermalized at $50^{\circ}$C and finally left to equilibrate on a rotating wheel turning at 10 rpm for one week.

Rheological experiments are performed with a stress-controlled rheometer (Anton Paar MCR $300$) equipped with a cone and plate geometry with a cone angle of $1^{\circ}$. Cone and plate are both roughened to minimize wall slip. The temperature is fixed to a precision of $0.1^{\circ}$C by a Peltier element connected to the bottom plate and a Peltier hood enclosing the sample environment. Flow-curves are determined at $10^{\circ}$C, $20^{\circ}$C and $50^{\circ}$C by decreasing progressively the applied shear rate and increasing logarithmically the measurement time per point to ensure steady flow conditions.

Flow cessation experiments are performed  at $10^{\circ}$C, $20^{\circ}$C and $50^{\circ}$C for shear rates of multiples of $1\times, 3\times, 6\times$, of $10^{-3}, 10^{-2}, 10^{-1}, 10^{0}, 10^{1}, 10^{2}$ s$^{-1}$. Prior to any flow cessation experiment we prepare the sample in oscillatory shear at a frequency of $1$Hz by performing an amplitude sweep, in which the strain amplitude is logarithmically decreased from $200$\% to $0.2$\% imposing 10 points per decade and a $10$s equilibration time per point. This range of strain successively covers the non-linear and linear range of elasticity, such that we presume that internal stresses are effectively minimized at the end of our preparation protocol \cite{lidon2017power}.
The results of the preparation experiment are also used to assess possible changes of the sample characteristics during long experiments, where we disregard experiments of an experimental series when the moduli have varied by more than 5\% from the beginning of the series. 
Following the preparation protocol, the shear rate of interest is applied till the stress reaches steady state. At this point the shear rate is set to zero, and the time dependence of the stress is recorded while the overall strain is maintained fixed. Because of the inertia of the tool, $\dot{\gamma} = 0$ is only reached after a given time delay, a time delay that depends on the shear rate applied prior to imposing the cessation command on the rheometer. Taking this experimental limitation into account, we restrict the analysis of the stress relaxations to the time beyond this time delay.\\

\textbf{Simulations:}
We consider an athermal jammed suspension of volume fraction $\phi = 70\%$, consisting of $\approx 10^5$ particles. Particles interact via shifted and truncated Lennard-Jones potentials \cite{weeks1971role}, 
\begin{eqnarray}
 U(r) &=& 4\epsilon \left[ \left( \frac{\sigma_{\alpha\beta}}{r}\right)^{12} - \left( \frac{\sigma_{\alpha\beta}}{r}\right)^6 \right] + \epsilon, r < r_c
\end{eqnarray}
 where $r_c = 2^{1/6} \sigma_{\alpha\beta}$ and $\epsilon = 1.0$ and $U(r)=0$ for $r \geq r_c$. The diameters $\sigma$ of the particles are drawn from a Gaussian distribution with a variance of 10\%, whose mean is used as unit length $a$.

 To prepare the jammed system of interest we first completely melt the FCC crystals at a temperature $T=5 \epsilon/k_B$. Thereafter, the system is allowed to equilibrate at that temperature using NVT molecular dynamics (MD) and subsequently quenched in steps with a cooling rate of $C_{\text{r}} = 5 \times 10^{-4}
\epsilon/(k_B  \uptau)$ until the final temperature of $0.001 \epsilon/k_B$ is reached. The cooling rate $C_{\text{r}}$ is here defined as $\Delta T/\Delta t$, where at each cooling step, the temperature is reduced by $\Delta T = 5 \times 10^{-3} \epsilon/k_B$ and the system relaxed for a duration of $\Delta t = 10 \uptau$, that is for $10^4$ MD steps with a MD time of $dt = 0.001 \uptau$, with  $\uptau = \sqrt{\frac{ma^2}{\epsilon}}$ the unit of time and m the unit of mass. Finally, we perform a conjugate gradient energy minimization protocol to obtain $T=0$ for the jammed system. 
 
 The jammed system is subjected to shear deformation at different imposed rates $\dot{\gamma}$ by performing dissipative particle dynamics (DPD) with changing box size $L(t) = L_0 (1+\dot{\gamma} dt)$ \cite{shrivastav2016heterogeneous,vasisht2020computational}. The DPD equation of motion is given by,
\begin{equation}
m\frac{d^2\vec{r}_i}{dt^2} = -(\nabla U(r_{ij}) + \zeta \sum_{j(\neq i)} (\hat{r}_{ij}.\vec{v}_{ij}))  \hat{r}_{ij}  
\end{equation}
where the first term is the conservative force and the second term is the dissipative force due to the solvent viscosity, with $\zeta$ is the damping coefficient, which is non-zero below the cutoff distance $2.5 a$. As $\uptau$ is not relevant to the physics explored in this work, we use the time unit $\uptau_o = \zeta a^2/\epsilon$ instead of $\uptau$, where $\uptau_o = 10 \uptau$ is the time needed for a particle experiencing the drag $\zeta$ to move under a unit force $\epsilon/a$ over a distance $a$.  Accordingly, the shear rate $\dot{\gamma}$ is expressed in the unit of $\uptau_o^{-1}$. The damping coefficients
$\zeta$ are chosen such that $m/\zeta$ is at least $0.1 \uptau$ to ensure that the simulations are performed at  overdamped conditions. For a detailed discussion of the simulation model, see Refs. \cite{shrivastav2016heterogeneous,vasisht2020computational}. We infer that the system has reached steady shear  flow when a linear velocity profile is reached. For the lower shear rates, where the time needed to reach steady flow is longer due to banding, we wait that the strain values are greater than $\gamma = 3.0$, even though the system appears to have reached a steady state at $\gamma \approx 0.4$ in the start up loading curves.  
%$X,Y,Z$ corresponds to flow, gradient and vorticity directions.
Once the system's kinetic energy is $\approx 10^{-20} \epsilon$, corresponding to a total force per particle of  $\approx 10^{-14} \epsilon/a$, is reached, we stop the flow by setting the shear rate to zero, fix the strain and allow the system to relax to a state that is in mechanical equilibrium. 
We determine the stress relaxation processes and the stress of the final state for preshear rates  ranging from $10^{-4}$ to $10^{0}  \uptau_o^{-1}$ and two damping coefficients $\zeta = 1$ and $5 \uptau_o \epsilon/a^2$. Simulations are performed with Lees-Edwards periodic boundary conditions using LAMMPS \cite{lammps}.

The shear stress ($\sigma$) is computed by using the Virial formula $\hat{\sigma} = \frac{1}{V} \sum_{i \neq j} \vec{r}_{ij} \otimes \vec{f}_{ij}$, where $V$ is the volume of the box, $\vec{r}_{ij}$ is the vector connecting the center of pairs $(i,j)$ and $\vec{f}_{ij}$ is the force between the pairs.

The mean squared displacement ($\langle \Delta r^2 \rangle$) is computed by $\langle \Delta r^2 \rangle = \langle \Delta x^2 + \Delta y^2 + \Delta z^2 \rangle$, where $\Delta r$ is the particle displacement for a given time interval and the average is performed over all particles. The reference configurations used for computing the displacements during the flow cessation are the steady flow configurations at the point of flow cessation. For the steady shear case, $\Delta r$ is computed from the non-affine displacements, obtained by removing the affine contributions from the particle displacement.

A scalar order parameter ($S$) is computed by using the displacement unit vectors $\hat{\Delta r}$, to determine the  tensor ${\bf Q} = \frac{1}{2}\langle 3 \Delta r_k \Delta r_l - \delta_{kl}\rangle$, with  $\Delta r_k$ denoting the component of the displacement unit vector, and the angular brackets $\langle ... \rangle$ signifying the average taken over all particle displacement unit vectors.  The  largest positive eigenvalue of ${\bf Q}$ is $S$; it is a measure of the degree of directionality in the particle displacement, with $S=1$ corresponding to all unit vectors aligned in one direction, while $S=0$ corresponds to random orientations of the unit vector. 

To further quantify correlations in the particle displacements, we compute the spatial correlation function \cite{weeks2007short}:
\begin{equation}\label{C_FLU}
C_{FLU} = \frac{\langle \delta u_i \delta u_j \rangle}{\langle \delta u^2 \rangle}
\end{equation}
where $\delta u_i = \Delta r_i - \langle \Delta r \rangle$,  $\Delta r_i$ is the magnitude of the displacement of particle $i$ and $\langle \Delta r \rangle$ is the averaged displacement. This analysis is done for dynamics observed upon flow cessation and for non-affine dynamics observed during steady flow. In both cases we observe that the dynamics at short times scales as  $\langle \Delta r^2 \rangle \propto t^2 $, reminiscent of ballistic motion.  The analysis of the particle displacements is restricted to the time window corresponding to the time window over which we observe quasi-ballistic motion in both experiments. 
%we compute the correlation functions in the ballistic regime and with the same reference configuration in the steady state.%
The dynamical correlation length is then determined from the decay rate of the correlation functions.

The structural evolution during flow cessation is studied by performing Voronoi tessellation on the configurations using VORO++ library \cite{rycroft2009voro++}. From this analysis, we obtain the statistics of Voronoi polyhedrons, and in particular, the amount of icosahedrons with $12$ faces, $30$ edges and $20$ vertices, which we quantify by $F_{ICO}$, the fraction of icosahedral configurations within the system.    
Icosahedron particle arrangements correspond to locally dense, overconstrained configurations, 
hence representing stiffer regions in the packing. This is supported by previous studies \cite{vasisht2020emergence,vasisht2020computational} denoting that the stability of well-annealed samples are due to the larger fractions of icosahedral regions and that these regions influence the stress overshoot and fluidization time of jammed suspensions.  
% Different measurables are averaged over atleast three independent samples for different shear rates. 
\\

%\backmatter

%\bmhead{Supplementary information}

%If your article has accompanying supplementary file/s please state so here. 

\textbf{Acknowledgments:}
The authors thank Gavin J. Donley and François Lavergne for fruitful discussions. M.M. and V.T. gratefully acknowledge financial support of the Swiss National Science Foundation (grant number 197340). Financial support from the National Science Foundation under grants nos. NSF DMR-2026842 (H.A.V. and E.D.G.) and NSF DMREF CBET—2118962 (E.D.G) is gratefully acknowledged.    

%\section*{Declarations}

%\nocite{*}
\bibliography{memory_FC_v1.bib}

\begin{thebibliography}{43}%
\makeatletter
\providecommand \@ifxundefined [1]{%
 \@ifx{#1\undefined}
}%
\providecommand \@ifnum [1]{%
 \ifnum #1\expandafter \@firstoftwo
 \else \expandafter \@secondoftwo
 \fi
}%
\providecommand \@ifx [1]{%
 \ifx #1\expandafter \@firstoftwo
 \else \expandafter \@secondoftwo
 \fi
}%
\providecommand \natexlab [1]{#1}%
\providecommand \enquote  [1]{``#1''}%
\providecommand \bibnamefont  [1]{#1}%
\providecommand \bibfnamefont [1]{#1}%
\providecommand \citenamefont [1]{#1}%
\providecommand \href@noop [0]{\@secondoftwo}%
\providecommand \href [0]{\begingroup \@sanitize@url \@href}%
\providecommand \@href[1]{\@@startlink{#1}\@@href}%
\providecommand \@@href[1]{\endgroup#1\@@endlink}%
\providecommand \@sanitize@url [0]{\catcode `\\12\catcode `\$12\catcode `\&12\catcode `\#12\catcode `\^12\catcode `\_12\catcode `\%12\relax}%
\providecommand \@@startlink[1]{}%
\providecommand \@@endlink[0]{}%
\providecommand \url  [0]{\begingroup\@sanitize@url \@url }%
\providecommand \@url [1]{\endgroup\@href {#1}{\urlprefix }}%
\providecommand \urlprefix  [0]{URL }%
\providecommand \Eprint [0]{\href }%
\providecommand \doibase [0]{http://dx.doi.org/}%
\providecommand \selectlanguage [0]{\@gobble}%
\providecommand \bibinfo  [0]{\@secondoftwo}%
\providecommand \bibfield  [0]{\@secondoftwo}%
\providecommand \translation [1]{[#1]}%
\providecommand \BibitemOpen [0]{}%
\providecommand \bibitemStop [0]{}%
\providecommand \bibitemNoStop [0]{.\EOS\space}%
\providecommand \EOS [0]{\spacefactor3000\relax}%
\providecommand \BibitemShut  [1]{\csname bibitem#1\endcsname}%
\let\auto@bib@innerbib\@empty
%</preamble>
\bibitem [{\citenamefont {Nguyen}\ and\ \citenamefont {Boger}(1992)}]{nguyen1992measuring}%
  \BibitemOpen
  \bibfield  {author} {\bibinfo {author} {\bibfnamefont {Q.}~\bibnamefont {Nguyen}}\ and\ \bibinfo {author} {\bibfnamefont {D.}~\bibnamefont {Boger}},\ }\href@noop {} {\bibfield  {journal} {\bibinfo  {journal} {Annual Review of Fluid Mechanics}\ }\textbf {\bibinfo {volume} {24}},\ \bibinfo {pages} {47} (\bibinfo {year} {1992})}\BibitemShut {NoStop}%
\bibitem [{\citenamefont {Bonn}\ \emph {et~al.}(2017)\citenamefont {Bonn}, \citenamefont {Denn}, \citenamefont {Berthier}, \citenamefont {Divoux},\ and\ \citenamefont {Manneville}}]{bonn2017yield}%
  \BibitemOpen
  \bibfield  {author} {\bibinfo {author} {\bibfnamefont {D.}~\bibnamefont {Bonn}}, \bibinfo {author} {\bibfnamefont {M.~M.}\ \bibnamefont {Denn}}, \bibinfo {author} {\bibfnamefont {L.}~\bibnamefont {Berthier}}, \bibinfo {author} {\bibfnamefont {T.}~\bibnamefont {Divoux}}, \ and\ \bibinfo {author} {\bibfnamefont {S.}~\bibnamefont {Manneville}},\ }\href@noop {} {\bibfield  {journal} {\bibinfo  {journal} {Reviews of Modern Physics}\ }\textbf {\bibinfo {volume} {89}},\ \bibinfo {pages} {035005} (\bibinfo {year} {2017})}\BibitemShut {NoStop}%
\bibitem [{\citenamefont {Nelson}\ \emph {et~al.}(2019)\citenamefont {Nelson}, \citenamefont {Schweizer}, \citenamefont {Rauzan}, \citenamefont {Nuzzo}, \citenamefont {Vermant},\ and\ \citenamefont {Ewoldt}}]{nelson2019designing}%
  \BibitemOpen
  \bibfield  {author} {\bibinfo {author} {\bibfnamefont {A.~Z.}\ \bibnamefont {Nelson}}, \bibinfo {author} {\bibfnamefont {K.~S.}\ \bibnamefont {Schweizer}}, \bibinfo {author} {\bibfnamefont {B.~M.}\ \bibnamefont {Rauzan}}, \bibinfo {author} {\bibfnamefont {R.~G.}\ \bibnamefont {Nuzzo}}, \bibinfo {author} {\bibfnamefont {J.}~\bibnamefont {Vermant}}, \ and\ \bibinfo {author} {\bibfnamefont {R.~H.}\ \bibnamefont {Ewoldt}},\ }\href@noop {} {\bibfield  {journal} {\bibinfo  {journal} {Current Opinion in Solid State and Materials Science}\ }\textbf {\bibinfo {volume} {23}},\ \bibinfo {pages} {100758} (\bibinfo {year} {2019})}\BibitemShut {NoStop}%
\bibitem [{\citenamefont {Nelson}\ \emph {et~al.}(2020)\citenamefont {Nelson}, \citenamefont {Kundukad}, \citenamefont {Wong}, \citenamefont {Khan},\ and\ \citenamefont {Doyle}}]{nelson2020embedded}%
  \BibitemOpen
  \bibfield  {author} {\bibinfo {author} {\bibfnamefont {A.~Z.}\ \bibnamefont {Nelson}}, \bibinfo {author} {\bibfnamefont {B.}~\bibnamefont {Kundukad}}, \bibinfo {author} {\bibfnamefont {W.~K.}\ \bibnamefont {Wong}}, \bibinfo {author} {\bibfnamefont {S.~A.}\ \bibnamefont {Khan}}, \ and\ \bibinfo {author} {\bibfnamefont {P.~S.}\ \bibnamefont {Doyle}},\ }\href@noop {} {\bibfield  {journal} {\bibinfo  {journal} {Proceedings of the National Academy of Sciences}\ }\textbf {\bibinfo {volume} {117}},\ \bibinfo {pages} {5671} (\bibinfo {year} {2020})}\BibitemShut {NoStop}%
\bibitem [{\citenamefont {Mohan}\ \emph {et~al.}(2015)\citenamefont {Mohan}, \citenamefont {Cloitre},\ and\ \citenamefont {Bonnecaze}}]{mohan2015build}%
  \BibitemOpen
  \bibfield  {author} {\bibinfo {author} {\bibfnamefont {L.}~\bibnamefont {Mohan}}, \bibinfo {author} {\bibfnamefont {M.}~\bibnamefont {Cloitre}}, \ and\ \bibinfo {author} {\bibfnamefont {R.~T.}\ \bibnamefont {Bonnecaze}},\ }\href@noop {} {\bibfield  {journal} {\bibinfo  {journal} {Journal of Rheology}\ }\textbf {\bibinfo {volume} {59}},\ \bibinfo {pages} {63} (\bibinfo {year} {2015})}\BibitemShut {NoStop}%
\bibitem [{\citenamefont {Hendricks}\ \emph {et~al.}(2019)\citenamefont {Hendricks}, \citenamefont {Louhichi}, \citenamefont {Metri}, \citenamefont {Fournier}, \citenamefont {Reddy}, \citenamefont {Bouteiller}, \citenamefont {Cloitre}, \citenamefont {Clasen}, \citenamefont {Vlassopoulos},\ and\ \citenamefont {Briels}}]{hendricks2019nonmonotonic}%
  \BibitemOpen
  \bibfield  {author} {\bibinfo {author} {\bibfnamefont {J.}~\bibnamefont {Hendricks}}, \bibinfo {author} {\bibfnamefont {A.}~\bibnamefont {Louhichi}}, \bibinfo {author} {\bibfnamefont {V.}~\bibnamefont {Metri}}, \bibinfo {author} {\bibfnamefont {R.}~\bibnamefont {Fournier}}, \bibinfo {author} {\bibfnamefont {N.}~\bibnamefont {Reddy}}, \bibinfo {author} {\bibfnamefont {L.}~\bibnamefont {Bouteiller}}, \bibinfo {author} {\bibfnamefont {M.}~\bibnamefont {Cloitre}}, \bibinfo {author} {\bibfnamefont {C.}~\bibnamefont {Clasen}}, \bibinfo {author} {\bibfnamefont {D.}~\bibnamefont {Vlassopoulos}}, \ and\ \bibinfo {author} {\bibfnamefont {W.}~\bibnamefont {Briels}},\ }\href@noop {} {\bibfield  {journal} {\bibinfo  {journal} {Physical review letters}\ }\textbf {\bibinfo {volume} {123}},\ \bibinfo {pages} {218003} (\bibinfo {year} {2019})}\BibitemShut {NoStop}%
\bibitem [{\citenamefont {Sudreau}\ \emph {et~al.}(2022{\natexlab{a}})\citenamefont {Sudreau}, \citenamefont {Auxois}, \citenamefont {Servel}, \citenamefont {L{\'e}colier}, \citenamefont {Manneville},\ and\ \citenamefont {Divoux}}]{sudreau2022residual}%
  \BibitemOpen
  \bibfield  {author} {\bibinfo {author} {\bibfnamefont {I.}~\bibnamefont {Sudreau}}, \bibinfo {author} {\bibfnamefont {M.}~\bibnamefont {Auxois}}, \bibinfo {author} {\bibfnamefont {M.}~\bibnamefont {Servel}}, \bibinfo {author} {\bibfnamefont {{\'E}.}~\bibnamefont {L{\'e}colier}}, \bibinfo {author} {\bibfnamefont {S.}~\bibnamefont {Manneville}}, \ and\ \bibinfo {author} {\bibfnamefont {T.}~\bibnamefont {Divoux}},\ }\href@noop {} {\bibfield  {journal} {\bibinfo  {journal} {Physical Review Materials}\ }\textbf {\bibinfo {volume} {6}},\ \bibinfo {pages} {L042601} (\bibinfo {year} {2022}{\natexlab{a}})}\BibitemShut {NoStop}%
\bibitem [{\citenamefont {Vasisht}\ \emph {et~al.}(2022)\citenamefont {Vasisht}, \citenamefont {Chaudhuri},\ and\ \citenamefont {Martens}}]{vasisht2022residual}%
  \BibitemOpen
  \bibfield  {author} {\bibinfo {author} {\bibfnamefont {V.~V.}\ \bibnamefont {Vasisht}}, \bibinfo {author} {\bibfnamefont {P.}~\bibnamefont {Chaudhuri}}, \ and\ \bibinfo {author} {\bibfnamefont {K.}~\bibnamefont {Martens}},\ }\href@noop {} {\bibfield  {journal} {\bibinfo  {journal} {Soft Matter}\ }\textbf {\bibinfo {volume} {18}},\ \bibinfo {pages} {6426} (\bibinfo {year} {2022})}\BibitemShut {NoStop}%
\bibitem [{\citenamefont {Sudreau}\ \emph {et~al.}(2022{\natexlab{b}})\citenamefont {Sudreau}, \citenamefont {Manneville}, \citenamefont {Servel},\ and\ \citenamefont {Divoux}}]{sudreau2022shear}%
  \BibitemOpen
  \bibfield  {author} {\bibinfo {author} {\bibfnamefont {I.}~\bibnamefont {Sudreau}}, \bibinfo {author} {\bibfnamefont {S.}~\bibnamefont {Manneville}}, \bibinfo {author} {\bibfnamefont {M.}~\bibnamefont {Servel}}, \ and\ \bibinfo {author} {\bibfnamefont {T.}~\bibnamefont {Divoux}},\ }\href@noop {} {\bibfield  {journal} {\bibinfo  {journal} {Journal of Rheology}\ }\textbf {\bibinfo {volume} {66}},\ \bibinfo {pages} {91} (\bibinfo {year} {2022}{\natexlab{b}})}\BibitemShut {NoStop}%
\bibitem [{\citenamefont {Nicolas}\ \emph {et~al.}(2018)\citenamefont {Nicolas}, \citenamefont {Ferrero}, \citenamefont {Martens},\ and\ \citenamefont {Barrat}}]{nicolas2018deformation}%
  \BibitemOpen
  \bibfield  {author} {\bibinfo {author} {\bibfnamefont {A.}~\bibnamefont {Nicolas}}, \bibinfo {author} {\bibfnamefont {E.~E.}\ \bibnamefont {Ferrero}}, \bibinfo {author} {\bibfnamefont {K.}~\bibnamefont {Martens}}, \ and\ \bibinfo {author} {\bibfnamefont {J.-L.}\ \bibnamefont {Barrat}},\ }\href@noop {} {\bibfield  {journal} {\bibinfo  {journal} {Reviews of Modern Physics}\ }\textbf {\bibinfo {volume} {90}},\ \bibinfo {pages} {045006} (\bibinfo {year} {2018})}\BibitemShut {NoStop}%
\bibitem [{\citenamefont {Vasisht}\ \emph {et~al.}(2018)\citenamefont {Vasisht}, \citenamefont {Dutta}, \citenamefont {Del~Gado},\ and\ \citenamefont {Blair}}]{vasisht2018rate}%
  \BibitemOpen
  \bibfield  {author} {\bibinfo {author} {\bibfnamefont {V.~V.}\ \bibnamefont {Vasisht}}, \bibinfo {author} {\bibfnamefont {S.~K.}\ \bibnamefont {Dutta}}, \bibinfo {author} {\bibfnamefont {E.}~\bibnamefont {Del~Gado}}, \ and\ \bibinfo {author} {\bibfnamefont {D.~L.}\ \bibnamefont {Blair}},\ }\href@noop {} {\bibfield  {journal} {\bibinfo  {journal} {Physical review letters}\ }\textbf {\bibinfo {volume} {120}},\ \bibinfo {pages} {018001} (\bibinfo {year} {2018})}\BibitemShut {NoStop}%
\bibitem [{\citenamefont {Khabaz}\ \emph {et~al.}(2020)\citenamefont {Khabaz}, \citenamefont {Cloitre},\ and\ \citenamefont {Bonnecaze}}]{khabaz2020particle}%
  \BibitemOpen
  \bibfield  {author} {\bibinfo {author} {\bibfnamefont {F.}~\bibnamefont {Khabaz}}, \bibinfo {author} {\bibfnamefont {M.}~\bibnamefont {Cloitre}}, \ and\ \bibinfo {author} {\bibfnamefont {R.~T.}\ \bibnamefont {Bonnecaze}},\ }\href@noop {} {\bibfield  {journal} {\bibinfo  {journal} {Journal of Rheology}\ }\textbf {\bibinfo {volume} {64}},\ \bibinfo {pages} {459} (\bibinfo {year} {2020})}\BibitemShut {NoStop}%
\bibitem [{\citenamefont {Aime}\ \emph {et~al.}(2023)\citenamefont {Aime}, \citenamefont {Truzzolillo}, \citenamefont {Pine}, \citenamefont {Ramos},\ and\ \citenamefont {Cipelletti}}]{aime2023unified}%
  \BibitemOpen
  \bibfield  {author} {\bibinfo {author} {\bibfnamefont {S.}~\bibnamefont {Aime}}, \bibinfo {author} {\bibfnamefont {D.}~\bibnamefont {Truzzolillo}}, \bibinfo {author} {\bibfnamefont {D.~J.}\ \bibnamefont {Pine}}, \bibinfo {author} {\bibfnamefont {L.}~\bibnamefont {Ramos}}, \ and\ \bibinfo {author} {\bibfnamefont {L.}~\bibnamefont {Cipelletti}},\ }\href@noop {} {\bibfield  {journal} {\bibinfo  {journal} {Nature Physics}\ ,\ \bibinfo {pages} {1}} (\bibinfo {year} {2023})}\BibitemShut {NoStop}%
\bibitem [{\citenamefont {Bandyopadhyay}\ \emph {et~al.}(2010)\citenamefont {Bandyopadhyay}, \citenamefont {Mohan},\ and\ \citenamefont {Joshi}}]{bandyopadhyay2010stress}%
  \BibitemOpen
  \bibfield  {author} {\bibinfo {author} {\bibfnamefont {R.}~\bibnamefont {Bandyopadhyay}}, \bibinfo {author} {\bibfnamefont {P.~H.}\ \bibnamefont {Mohan}}, \ and\ \bibinfo {author} {\bibfnamefont {Y.~M.}\ \bibnamefont {Joshi}},\ }\href@noop {} {\bibfield  {journal} {\bibinfo  {journal} {Soft Matter}\ }\textbf {\bibinfo {volume} {6}},\ \bibinfo {pages} {1462} (\bibinfo {year} {2010})}\BibitemShut {NoStop}%
\bibitem [{\citenamefont {Barik}\ and\ \citenamefont {Majumdar}(2022)}]{barik2022origin}%
  \BibitemOpen
  \bibfield  {author} {\bibinfo {author} {\bibfnamefont {S.}~\bibnamefont {Barik}}\ and\ \bibinfo {author} {\bibfnamefont {S.}~\bibnamefont {Majumdar}},\ }\href@noop {} {\bibfield  {journal} {\bibinfo  {journal} {Physical Review Letters}\ }\textbf {\bibinfo {volume} {128}},\ \bibinfo {pages} {258002} (\bibinfo {year} {2022})}\BibitemShut {NoStop}%
\bibitem [{\citenamefont {Ballauff}\ \emph {et~al.}(2013)\citenamefont {Ballauff}, \citenamefont {Brader}, \citenamefont {Egelhaaf}, \citenamefont {Fuchs}, \citenamefont {Horbach}, \citenamefont {Koumakis}, \citenamefont {Kr{\"u}ger}, \citenamefont {Laurati}, \citenamefont {Mutch}, \citenamefont {Petekidis} \emph {et~al.}}]{ballauff2013residual}%
  \BibitemOpen
  \bibfield  {author} {\bibinfo {author} {\bibfnamefont {M.}~\bibnamefont {Ballauff}}, \bibinfo {author} {\bibfnamefont {J.~M.}\ \bibnamefont {Brader}}, \bibinfo {author} {\bibfnamefont {S.~U.}\ \bibnamefont {Egelhaaf}}, \bibinfo {author} {\bibfnamefont {M.}~\bibnamefont {Fuchs}}, \bibinfo {author} {\bibfnamefont {J.}~\bibnamefont {Horbach}}, \bibinfo {author} {\bibfnamefont {N.}~\bibnamefont {Koumakis}}, \bibinfo {author} {\bibfnamefont {M.}~\bibnamefont {Kr{\"u}ger}}, \bibinfo {author} {\bibfnamefont {M.}~\bibnamefont {Laurati}}, \bibinfo {author} {\bibfnamefont {K.~J.}\ \bibnamefont {Mutch}}, \bibinfo {author} {\bibfnamefont {G.}~\bibnamefont {Petekidis}},  \emph {et~al.},\ }\href@noop {} {\bibfield  {journal} {\bibinfo  {journal} {Physical review letters}\ }\textbf {\bibinfo {volume} {110}},\ \bibinfo {pages} {215701} (\bibinfo {year} {2013})}\BibitemShut {NoStop}%
\bibitem [{\citenamefont {Bhattacharyya}\ \emph {et~al.}(2023)\citenamefont {Bhattacharyya}, \citenamefont {Jacob}, \citenamefont {Petekidis},\ and\ \citenamefont {Joshi}}]{bhattacharyya2023nature}%
  \BibitemOpen
  \bibfield  {author} {\bibinfo {author} {\bibfnamefont {T.}~\bibnamefont {Bhattacharyya}}, \bibinfo {author} {\bibfnamefont {A.~R.}\ \bibnamefont {Jacob}}, \bibinfo {author} {\bibfnamefont {G.}~\bibnamefont {Petekidis}}, \ and\ \bibinfo {author} {\bibfnamefont {Y.~M.}\ \bibnamefont {Joshi}},\ }\href@noop {} {\bibfield  {journal} {\bibinfo  {journal} {Journal of Rheology}\ }\textbf {\bibinfo {volume} {67}},\ \bibinfo {pages} {461} (\bibinfo {year} {2023})}\BibitemShut {NoStop}%
\bibitem [{\citenamefont {Caggioni}\ \emph {et~al.}(2020)\citenamefont {Caggioni}, \citenamefont {Trappe},\ and\ \citenamefont {Spicer}}]{caggioni2020variations}%
  \BibitemOpen
  \bibfield  {author} {\bibinfo {author} {\bibfnamefont {M.}~\bibnamefont {Caggioni}}, \bibinfo {author} {\bibfnamefont {V.}~\bibnamefont {Trappe}}, \ and\ \bibinfo {author} {\bibfnamefont {P.~T.}\ \bibnamefont {Spicer}},\ }\href@noop {} {\bibfield  {journal} {\bibinfo  {journal} {Journal of Rheology}\ }\textbf {\bibinfo {volume} {64}},\ \bibinfo {pages} {413} (\bibinfo {year} {2020})}\BibitemShut {NoStop}%
\bibitem [{\citenamefont {Song}\ \emph {et~al.}(2022)\citenamefont {Song}, \citenamefont {Zhang}, \citenamefont {de~Quesada}, \citenamefont {Rizvi}, \citenamefont {Tracy}, \citenamefont {Ilavsky}, \citenamefont {Narayanan}, \citenamefont {Del~Gado}, \citenamefont {Leheny}, \citenamefont {Holten-Andersen} \emph {et~al.}}]{song2022microscopic}%
  \BibitemOpen
  \bibfield  {author} {\bibinfo {author} {\bibfnamefont {J.}~\bibnamefont {Song}}, \bibinfo {author} {\bibfnamefont {Q.}~\bibnamefont {Zhang}}, \bibinfo {author} {\bibfnamefont {F.}~\bibnamefont {de~Quesada}}, \bibinfo {author} {\bibfnamefont {M.~H.}\ \bibnamefont {Rizvi}}, \bibinfo {author} {\bibfnamefont {J.~B.}\ \bibnamefont {Tracy}}, \bibinfo {author} {\bibfnamefont {J.}~\bibnamefont {Ilavsky}}, \bibinfo {author} {\bibfnamefont {S.}~\bibnamefont {Narayanan}}, \bibinfo {author} {\bibfnamefont {E.}~\bibnamefont {Del~Gado}}, \bibinfo {author} {\bibfnamefont {R.~L.}\ \bibnamefont {Leheny}}, \bibinfo {author} {\bibfnamefont {N.}~\bibnamefont {Holten-Andersen}},  \emph {et~al.},\ }\href@noop {} {\bibfield  {journal} {\bibinfo  {journal} {Proceedings of the National Academy of Sciences}\ }\textbf {\bibinfo {volume} {119}},\ \bibinfo {pages} {e2201566119} (\bibinfo {year} {2022})}\BibitemShut {NoStop}%
\bibitem [{\citenamefont {Hwang}\ \emph {et~al.}(2016)\citenamefont {Hwang}, \citenamefont {Riggleman},\ and\ \citenamefont {Crocker}}]{hwang2016understanding}%
  \BibitemOpen
  \bibfield  {author} {\bibinfo {author} {\bibfnamefont {H.~J.}\ \bibnamefont {Hwang}}, \bibinfo {author} {\bibfnamefont {R.~A.}\ \bibnamefont {Riggleman}}, \ and\ \bibinfo {author} {\bibfnamefont {J.~C.}\ \bibnamefont {Crocker}},\ }\href@noop {} {\bibfield  {journal} {\bibinfo  {journal} {Nature materials}\ }\textbf {\bibinfo {volume} {15}},\ \bibinfo {pages} {1031} (\bibinfo {year} {2016})}\BibitemShut {NoStop}%
\bibitem [{\citenamefont {Vasisht}\ and\ \citenamefont {Del~Gado}(2020)}]{vasisht2020computational}%
  \BibitemOpen
  \bibfield  {author} {\bibinfo {author} {\bibfnamefont {V.~V.}\ \bibnamefont {Vasisht}}\ and\ \bibinfo {author} {\bibfnamefont {E.}~\bibnamefont {Del~Gado}},\ }\href@noop {} {\bibfield  {journal} {\bibinfo  {journal} {Physical Review E}\ }\textbf {\bibinfo {volume} {102}},\ \bibinfo {pages} {012603} (\bibinfo {year} {2020})}\BibitemShut {NoStop}%
\bibitem [{\citenamefont {Vasisht}\ \emph {et~al.}(2020)\citenamefont {Vasisht}, \citenamefont {Roberts},\ and\ \citenamefont {Del~Gado}}]{vasisht2020emergence}%
  \BibitemOpen
  \bibfield  {author} {\bibinfo {author} {\bibfnamefont {V.~V.}\ \bibnamefont {Vasisht}}, \bibinfo {author} {\bibfnamefont {G.}~\bibnamefont {Roberts}}, \ and\ \bibinfo {author} {\bibfnamefont {E.}~\bibnamefont {Del~Gado}},\ }\href@noop {} {\bibfield  {journal} {\bibinfo  {journal} {Physical Review E}\ }\textbf {\bibinfo {volume} {102}},\ \bibinfo {pages} {010604} (\bibinfo {year} {2020})}\BibitemShut {NoStop}%
\bibitem [{\citenamefont {Pinney}\ \emph {et~al.}(2016)\citenamefont {Pinney}, \citenamefont {Liverpool},\ and\ \citenamefont {Royall}}]{pinney2016structure}%
  \BibitemOpen
  \bibfield  {author} {\bibinfo {author} {\bibfnamefont {R.}~\bibnamefont {Pinney}}, \bibinfo {author} {\bibfnamefont {T.~B.}\ \bibnamefont {Liverpool}}, \ and\ \bibinfo {author} {\bibfnamefont {C.~P.}\ \bibnamefont {Royall}},\ }\href@noop {} {\bibfield  {journal} {\bibinfo  {journal} {The Journal of Chemical Physics}\ }\textbf {\bibinfo {volume} {145}} (\bibinfo {year} {2016})}\BibitemShut {NoStop}%
\bibitem [{\citenamefont {Mohan}\ \emph {et~al.}(2013)\citenamefont {Mohan}, \citenamefont {Bonnecaze},\ and\ \citenamefont {Cloitre}}]{mohan2013microscopic}%
  \BibitemOpen
  \bibfield  {author} {\bibinfo {author} {\bibfnamefont {L.}~\bibnamefont {Mohan}}, \bibinfo {author} {\bibfnamefont {R.~T.}\ \bibnamefont {Bonnecaze}}, \ and\ \bibinfo {author} {\bibfnamefont {M.}~\bibnamefont {Cloitre}},\ }\href@noop {} {\bibfield  {journal} {\bibinfo  {journal} {Physical Review Letters}\ }\textbf {\bibinfo {volume} {111}},\ \bibinfo {pages} {268301} (\bibinfo {year} {2013})}\BibitemShut {NoStop}%
\bibitem [{\citenamefont {Cuny}\ \emph {et~al.}(2021)\citenamefont {Cuny}, \citenamefont {Mari},\ and\ \citenamefont {Bertin}}]{cuny2021microscopic}%
  \BibitemOpen
  \bibfield  {author} {\bibinfo {author} {\bibfnamefont {N.}~\bibnamefont {Cuny}}, \bibinfo {author} {\bibfnamefont {R.}~\bibnamefont {Mari}}, \ and\ \bibinfo {author} {\bibfnamefont {E.}~\bibnamefont {Bertin}},\ }\href@noop {} {\bibfield  {journal} {\bibinfo  {journal} {Physical Review Letters}\ }\textbf {\bibinfo {volume} {127}},\ \bibinfo {pages} {218003} (\bibinfo {year} {2021})}\BibitemShut {NoStop}%
\bibitem [{\citenamefont {Chaikin}\ \emph {et~al.}(1995)\citenamefont {Chaikin}, \citenamefont {Lubensky},\ and\ \citenamefont {Witten}}]{chaikin1995principles}%
  \BibitemOpen
  \bibfield  {author} {\bibinfo {author} {\bibfnamefont {P.~M.}\ \bibnamefont {Chaikin}}, \bibinfo {author} {\bibfnamefont {T.~C.}\ \bibnamefont {Lubensky}}, \ and\ \bibinfo {author} {\bibfnamefont {T.~A.}\ \bibnamefont {Witten}},\ }\href@noop {} {\emph {\bibinfo {title} {Principles of condensed matter physics}}},\ Vol.~\bibinfo {volume} {10}\ (\bibinfo  {publisher} {Cambridge university press Cambridge},\ \bibinfo {year} {1995})\BibitemShut {NoStop}%
\bibitem [{\citenamefont {Picard}\ \emph {et~al.}(2002)\citenamefont {Picard}, \citenamefont {Ajdari}, \citenamefont {Bocquet},\ and\ \citenamefont {Lequeux}}]{picard2002simple}%
  \BibitemOpen
  \bibfield  {author} {\bibinfo {author} {\bibfnamefont {G.}~\bibnamefont {Picard}}, \bibinfo {author} {\bibfnamefont {A.}~\bibnamefont {Ajdari}}, \bibinfo {author} {\bibfnamefont {L.}~\bibnamefont {Bocquet}}, \ and\ \bibinfo {author} {\bibfnamefont {F.}~\bibnamefont {Lequeux}},\ }\href@noop {} {\bibfield  {journal} {\bibinfo  {journal} {Physical Review E}\ }\textbf {\bibinfo {volume} {66}},\ \bibinfo {pages} {051501} (\bibinfo {year} {2002})}\BibitemShut {NoStop}%
\bibitem [{\citenamefont {Bocquet}\ \emph {et~al.}(2009)\citenamefont {Bocquet}, \citenamefont {Colin},\ and\ \citenamefont {Ajdari}}]{bocquet2009kinetic}%
  \BibitemOpen
  \bibfield  {author} {\bibinfo {author} {\bibfnamefont {L.}~\bibnamefont {Bocquet}}, \bibinfo {author} {\bibfnamefont {A.}~\bibnamefont {Colin}}, \ and\ \bibinfo {author} {\bibfnamefont {A.}~\bibnamefont {Ajdari}},\ }\href@noop {} {\bibfield  {journal} {\bibinfo  {journal} {Physical review letters}\ }\textbf {\bibinfo {volume} {103}},\ \bibinfo {pages} {036001} (\bibinfo {year} {2009})}\BibitemShut {NoStop}%
\bibitem [{\citenamefont {Ferrero}\ \emph {et~al.}(2014)\citenamefont {Ferrero}, \citenamefont {Martens},\ and\ \citenamefont {Barrat}}]{ferrero2014relaxation}%
  \BibitemOpen
  \bibfield  {author} {\bibinfo {author} {\bibfnamefont {E.~E.}\ \bibnamefont {Ferrero}}, \bibinfo {author} {\bibfnamefont {K.}~\bibnamefont {Martens}}, \ and\ \bibinfo {author} {\bibfnamefont {J.-L.}\ \bibnamefont {Barrat}},\ }\href@noop {} {\bibfield  {journal} {\bibinfo  {journal} {Physical review letters}\ }\textbf {\bibinfo {volume} {113}},\ \bibinfo {pages} {248301} (\bibinfo {year} {2014})}\BibitemShut {NoStop}%
\bibitem [{\citenamefont {Lin}\ and\ \citenamefont {Wyart}(2018)}]{lin2018microscopic}%
  \BibitemOpen
  \bibfield  {author} {\bibinfo {author} {\bibfnamefont {J.}~\bibnamefont {Lin}}\ and\ \bibinfo {author} {\bibfnamefont {M.}~\bibnamefont {Wyart}},\ }\href@noop {} {\bibfield  {journal} {\bibinfo  {journal} {Physical review E}\ }\textbf {\bibinfo {volume} {97}},\ \bibinfo {pages} {012603} (\bibinfo {year} {2018})}\BibitemShut {NoStop}%
\bibitem [{\citenamefont {Benzi}\ \emph {et~al.}(2021)\citenamefont {Benzi}, \citenamefont {Divoux}, \citenamefont {Barentin}, \citenamefont {Manneville}, \citenamefont {Sbragaglia},\ and\ \citenamefont {Toschi}}]{benzi2021continuum}%
  \BibitemOpen
  \bibfield  {author} {\bibinfo {author} {\bibfnamefont {R.}~\bibnamefont {Benzi}}, \bibinfo {author} {\bibfnamefont {T.}~\bibnamefont {Divoux}}, \bibinfo {author} {\bibfnamefont {C.}~\bibnamefont {Barentin}}, \bibinfo {author} {\bibfnamefont {S.}~\bibnamefont {Manneville}}, \bibinfo {author} {\bibfnamefont {M.}~\bibnamefont {Sbragaglia}}, \ and\ \bibinfo {author} {\bibfnamefont {F.}~\bibnamefont {Toschi}},\ }\href@noop {} {\bibfield  {journal} {\bibinfo  {journal} {Physical Review E}\ }\textbf {\bibinfo {volume} {104}},\ \bibinfo {pages} {034612} (\bibinfo {year} {2021})}\BibitemShut {NoStop}%
\bibitem [{\citenamefont {Clemmer}\ \emph {et~al.}(2021{\natexlab{a}})\citenamefont {Clemmer}, \citenamefont {Salerno},\ and\ \citenamefont {Robbins}}]{clemmer2021criticality}%
  \BibitemOpen
  \bibfield  {author} {\bibinfo {author} {\bibfnamefont {J.~T.}\ \bibnamefont {Clemmer}}, \bibinfo {author} {\bibfnamefont {K.~M.}\ \bibnamefont {Salerno}}, \ and\ \bibinfo {author} {\bibfnamefont {M.~O.}\ \bibnamefont {Robbins}},\ }\href@noop {} {\bibfield  {journal} {\bibinfo  {journal} {Physical Review E}\ }\textbf {\bibinfo {volume} {103}},\ \bibinfo {pages} {042606} (\bibinfo {year} {2021}{\natexlab{a}})}\BibitemShut {NoStop}%
\bibitem [{\citenamefont {Clemmer}\ \emph {et~al.}(2021{\natexlab{b}})\citenamefont {Clemmer}, \citenamefont {Salerno},\ and\ \citenamefont {Robbins}}]{clemmer2021criticality1}%
  \BibitemOpen
  \bibfield  {author} {\bibinfo {author} {\bibfnamefont {J.~T.}\ \bibnamefont {Clemmer}}, \bibinfo {author} {\bibfnamefont {K.~M.}\ \bibnamefont {Salerno}}, \ and\ \bibinfo {author} {\bibfnamefont {M.~O.}\ \bibnamefont {Robbins}},\ }\href@noop {} {\bibfield  {journal} {\bibinfo  {journal} {Physical Review E}\ }\textbf {\bibinfo {volume} {103}},\ \bibinfo {pages} {042605} (\bibinfo {year} {2021}{\natexlab{b}})}\BibitemShut {NoStop}%
\bibitem [{\citenamefont {Lin}\ \emph {et~al.}(2014)\citenamefont {Lin}, \citenamefont {Lerner}, \citenamefont {Rosso},\ and\ \citenamefont {Wyart}}]{lin2014scaling}%
  \BibitemOpen
  \bibfield  {author} {\bibinfo {author} {\bibfnamefont {J.}~\bibnamefont {Lin}}, \bibinfo {author} {\bibfnamefont {E.}~\bibnamefont {Lerner}}, \bibinfo {author} {\bibfnamefont {A.}~\bibnamefont {Rosso}}, \ and\ \bibinfo {author} {\bibfnamefont {M.}~\bibnamefont {Wyart}},\ }\href@noop {} {\bibfield  {journal} {\bibinfo  {journal} {Proceedings of the National Academy of Sciences}\ }\textbf {\bibinfo {volume} {111}},\ \bibinfo {pages} {14382} (\bibinfo {year} {2014})}\BibitemShut {NoStop}%
\bibitem [{\citenamefont {Cuny}\ \emph {et~al.}(2022)\citenamefont {Cuny}, \citenamefont {Bertin},\ and\ \citenamefont {Mari}}]{cuny2022dynamics}%
  \BibitemOpen
  \bibfield  {author} {\bibinfo {author} {\bibfnamefont {N.}~\bibnamefont {Cuny}}, \bibinfo {author} {\bibfnamefont {E.}~\bibnamefont {Bertin}}, \ and\ \bibinfo {author} {\bibfnamefont {R.}~\bibnamefont {Mari}},\ }\href@noop {} {\bibfield  {journal} {\bibinfo  {journal} {Soft Matter}\ }\textbf {\bibinfo {volume} {18}},\ \bibinfo {pages} {328} (\bibinfo {year} {2022})}\BibitemShut {NoStop}%
\bibitem [{\citenamefont {Conley}\ \emph {et~al.}(2017)\citenamefont {Conley}, \citenamefont {Aebischer}, \citenamefont {N{\"o}jd}, \citenamefont {Schurtenberger},\ and\ \citenamefont {Scheffold}}]{conley2017jamming}%
  \BibitemOpen
  \bibfield  {author} {\bibinfo {author} {\bibfnamefont {G.~M.}\ \bibnamefont {Conley}}, \bibinfo {author} {\bibfnamefont {P.}~\bibnamefont {Aebischer}}, \bibinfo {author} {\bibfnamefont {S.}~\bibnamefont {N{\"o}jd}}, \bibinfo {author} {\bibfnamefont {P.}~\bibnamefont {Schurtenberger}}, \ and\ \bibinfo {author} {\bibfnamefont {F.}~\bibnamefont {Scheffold}},\ }\href@noop {} {\bibfield  {journal} {\bibinfo  {journal} {Science advances}\ }\textbf {\bibinfo {volume} {3}},\ \bibinfo {pages} {e1700969} (\bibinfo {year} {2017})}\BibitemShut {NoStop}%
\bibitem [{\citenamefont {Conley}\ \emph {et~al.}(2019)\citenamefont {Conley}, \citenamefont {Zhang}, \citenamefont {Aebischer}, \citenamefont {Harden},\ and\ \citenamefont {Scheffold}}]{conley2019relationship}%
  \BibitemOpen
  \bibfield  {author} {\bibinfo {author} {\bibfnamefont {G.~M.}\ \bibnamefont {Conley}}, \bibinfo {author} {\bibfnamefont {C.}~\bibnamefont {Zhang}}, \bibinfo {author} {\bibfnamefont {P.}~\bibnamefont {Aebischer}}, \bibinfo {author} {\bibfnamefont {J.~L.}\ \bibnamefont {Harden}}, \ and\ \bibinfo {author} {\bibfnamefont {F.}~\bibnamefont {Scheffold}},\ }\href@noop {} {\bibfield  {journal} {\bibinfo  {journal} {Nature communications}\ }\textbf {\bibinfo {volume} {10}},\ \bibinfo {pages} {2436} (\bibinfo {year} {2019})}\BibitemShut {NoStop}%
\bibitem [{\citenamefont {Lidon}\ \emph {et~al.}(2017)\citenamefont {Lidon}, \citenamefont {Villa},\ and\ \citenamefont {Manneville}}]{lidon2017power}%
  \BibitemOpen
  \bibfield  {author} {\bibinfo {author} {\bibfnamefont {P.}~\bibnamefont {Lidon}}, \bibinfo {author} {\bibfnamefont {L.}~\bibnamefont {Villa}}, \ and\ \bibinfo {author} {\bibfnamefont {S.}~\bibnamefont {Manneville}},\ }\href {\doibase 10.1007/s00397-016-0961-4} {\bibfield  {journal} {\bibinfo  {journal} {Rheologica Acta}\ }\textbf {\bibinfo {volume} {56}},\ \bibinfo {pages} {307} (\bibinfo {year} {2017})}\BibitemShut {NoStop}%
\bibitem [{\citenamefont {Weeks}\ \emph {et~al.}(1971)\citenamefont {Weeks}, \citenamefont {Chandler},\ and\ \citenamefont {Andersen}}]{weeks1971role}%
  \BibitemOpen
  \bibfield  {author} {\bibinfo {author} {\bibfnamefont {J.~D.}\ \bibnamefont {Weeks}}, \bibinfo {author} {\bibfnamefont {D.}~\bibnamefont {Chandler}}, \ and\ \bibinfo {author} {\bibfnamefont {H.~C.}\ \bibnamefont {Andersen}},\ }\href@noop {} {\bibfield  {journal} {\bibinfo  {journal} {The Journal of chemical physics}\ }\textbf {\bibinfo {volume} {54}},\ \bibinfo {pages} {5237} (\bibinfo {year} {1971})}\BibitemShut {NoStop}%
\bibitem [{\citenamefont {Shrivastav}\ \emph {et~al.}(2016)\citenamefont {Shrivastav}, \citenamefont {Chaudhuri},\ and\ \citenamefont {Horbach}}]{shrivastav2016heterogeneous}%
  \BibitemOpen
  \bibfield  {author} {\bibinfo {author} {\bibfnamefont {G.~P.}\ \bibnamefont {Shrivastav}}, \bibinfo {author} {\bibfnamefont {P.}~\bibnamefont {Chaudhuri}}, \ and\ \bibinfo {author} {\bibfnamefont {J.}~\bibnamefont {Horbach}},\ }\href@noop {} {\bibfield  {journal} {\bibinfo  {journal} {Journal of Rheology}\ }\textbf {\bibinfo {volume} {60}},\ \bibinfo {pages} {835} (\bibinfo {year} {2016})}\BibitemShut {NoStop}%
\bibitem [{\citenamefont {Plimpton}(1995)}]{lammps}%
  \BibitemOpen
  \bibfield  {author} {\bibinfo {author} {\bibfnamefont {S.}~\bibnamefont {Plimpton}},\ }\href@noop {} {\bibfield  {journal} {\bibinfo  {journal} {Journal of computational physics}\ }\textbf {\bibinfo {volume} {117}},\ \bibinfo {pages} {1} (\bibinfo {year} {1995})}\BibitemShut {NoStop}%
\bibitem [{\citenamefont {Weeks}\ \emph {et~al.}(2007)\citenamefont {Weeks}, \citenamefont {Crocker},\ and\ \citenamefont {Weitz}}]{weeks2007short}%
  \BibitemOpen
  \bibfield  {author} {\bibinfo {author} {\bibfnamefont {E.~R.}\ \bibnamefont {Weeks}}, \bibinfo {author} {\bibfnamefont {J.~C.}\ \bibnamefont {Crocker}}, \ and\ \bibinfo {author} {\bibfnamefont {D.~A.}\ \bibnamefont {Weitz}},\ }\href@noop {} {\bibfield  {journal} {\bibinfo  {journal} {Journal of Physics: Condensed Matter}\ }\textbf {\bibinfo {volume} {19}},\ \bibinfo {pages} {205131} (\bibinfo {year} {2007})}\BibitemShut {NoStop}%
\bibitem [{\citenamefont {Rycroft}(2009)}]{rycroft2009voro++}%
  \BibitemOpen
  \bibfield  {author} {\bibinfo {author} {\bibfnamefont {C.}~\bibnamefont {Rycroft}},\ }\href@noop {} {\emph {\bibinfo {title} {Voro++: A three-dimensional Voronoi cell library in C++}}},\ \bibinfo {type} {Tech. Rep.}\ (\bibinfo  {institution} {Lawrence Berkeley National Lab.(LBNL), Berkeley, CA (United States)},\ \bibinfo {year} {2009})\BibitemShut {NoStop}%
\end{thebibliography}%

\end{document}

%\maketitle 
\begin{center}
%\textbf{Supplementary Information}
\textbf{Memory of shear flow in jammed suspensions (Supplementary Material)}
\\
H. A. Vinutha, Manon Marchand, Marco Caggioni, Vishwas V. Vasisht, Emanuela Del Gado, and Veronique Trappe
\date{\today}
\end{center}

%\begin{abstract}

%Here we present additional data and supplemental information regarding the analysis presented in the paper, namely: (i) Raw stress relaxation data of a model jammed suspension from experiments \& simulations. (ii) Particle displacements along different directions. (iii) Dynamical correlations for different shear rates at $\zeta = 5 \tau_o \epsilon/a^2$, different runs with the same shear rate at $\zeta = 1 \tau_o \epsilon/a^2$ and unscaled data of the correlation length.  (iv) Unscaled data of fraction of icosahedrons in steady state and the time evolution of $F_{ICO}$ for two different damping coefficients. (v) Estimation of the volume fraction of the experimental system. 
%Scaling of the correlation length as a function of the distance from yield stress. 

%\end{abstract}
%\begin{enumerate}

\section{Characteristics of stress relaxation upon flow cessation}
\begin{figure*}[h!]
%\centering
\includegraphics[scale=0.3]{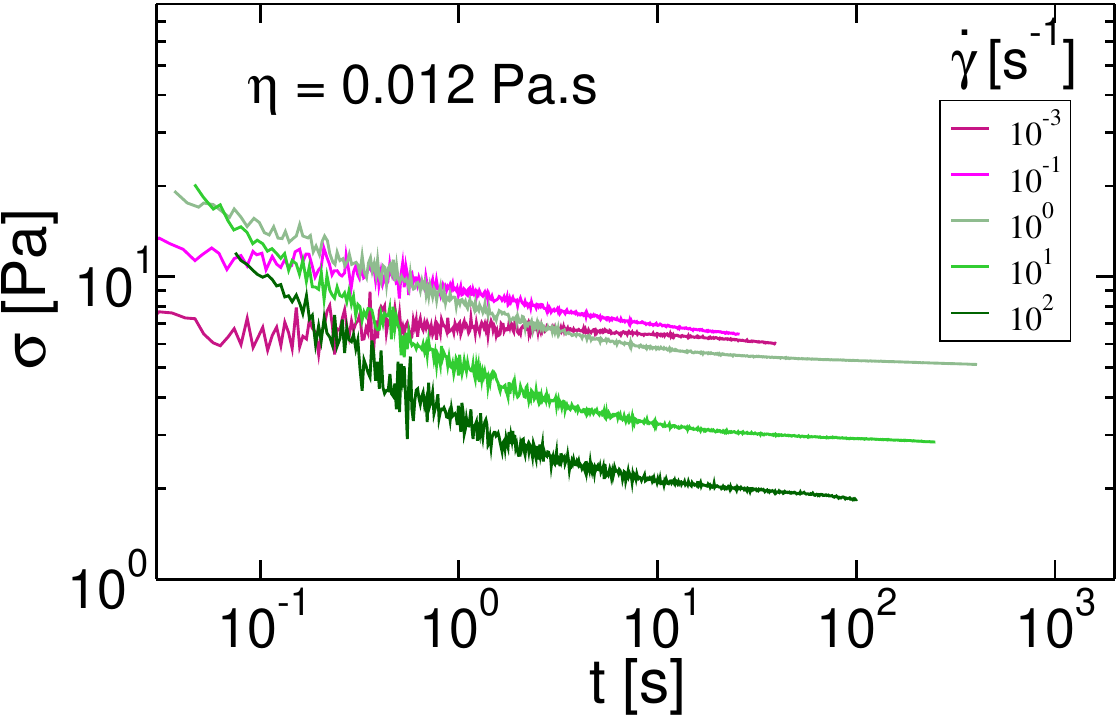}
\includegraphics[scale=0.3]{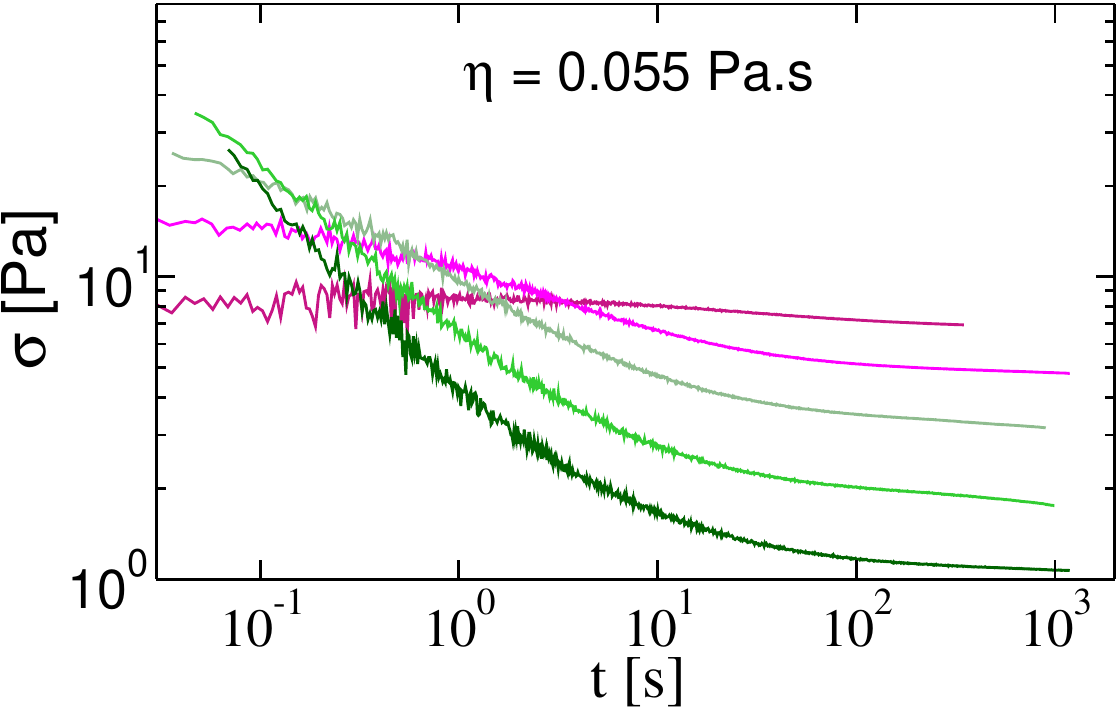}
\includegraphics[scale=0.29]{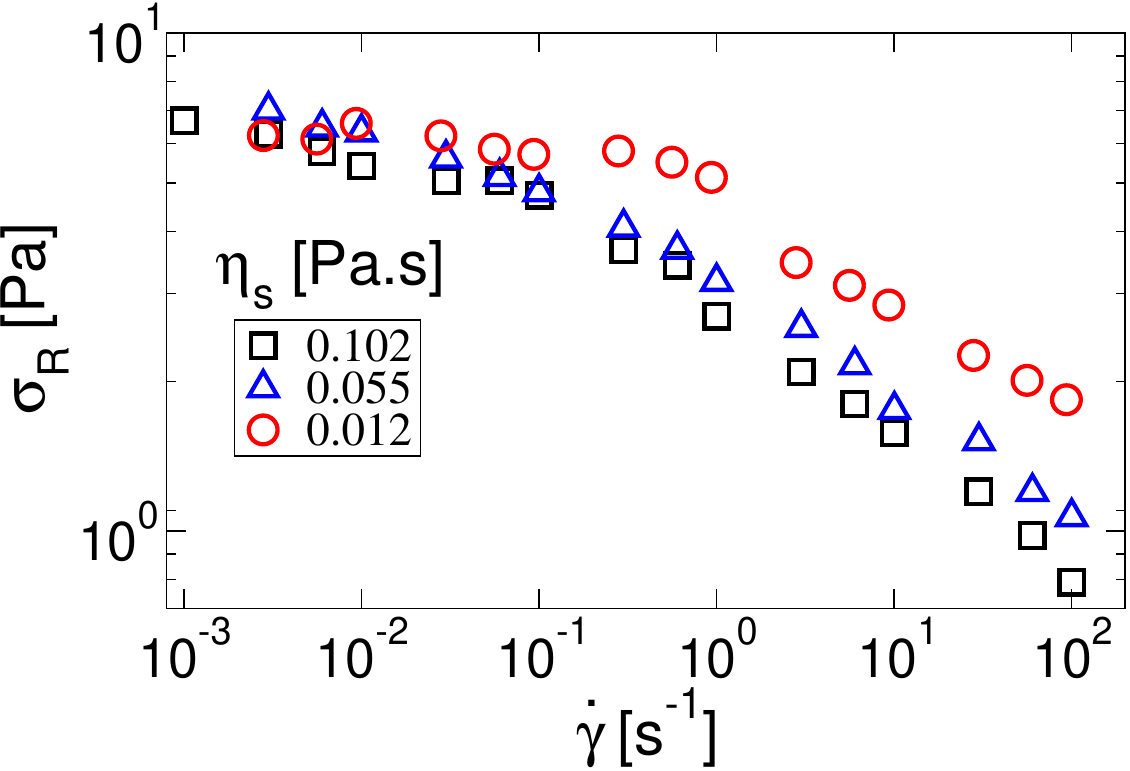}

\includegraphics[scale=0.3]{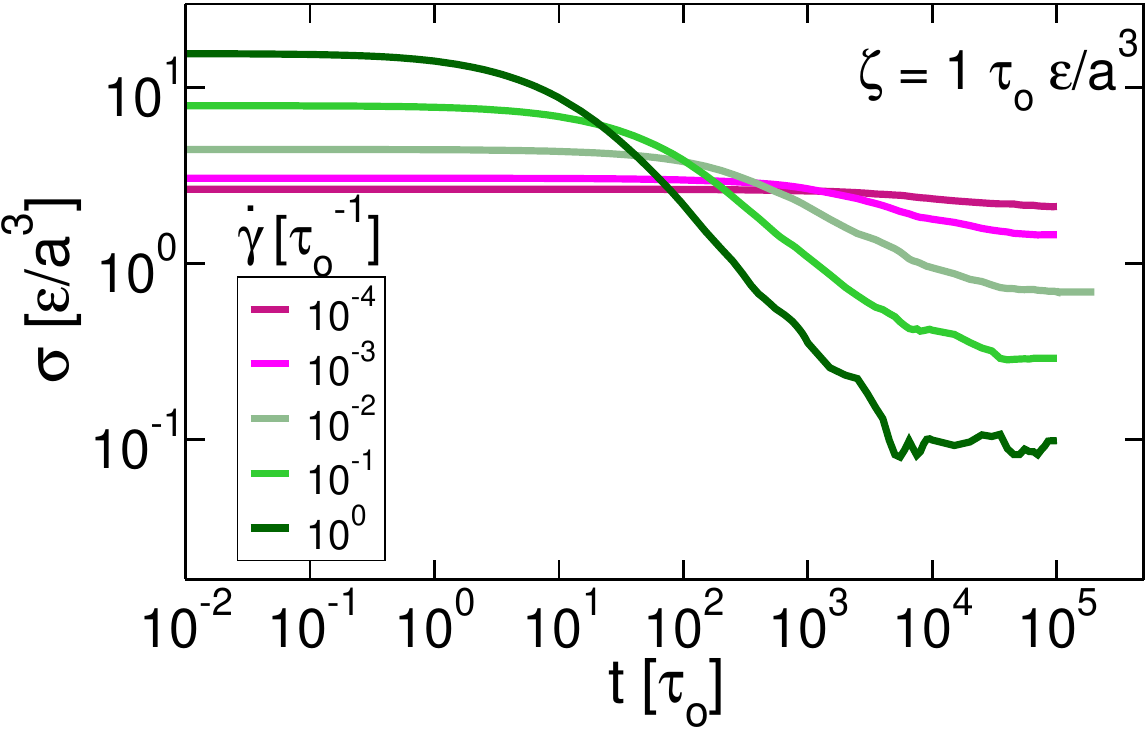}
\includegraphics[scale=0.29]{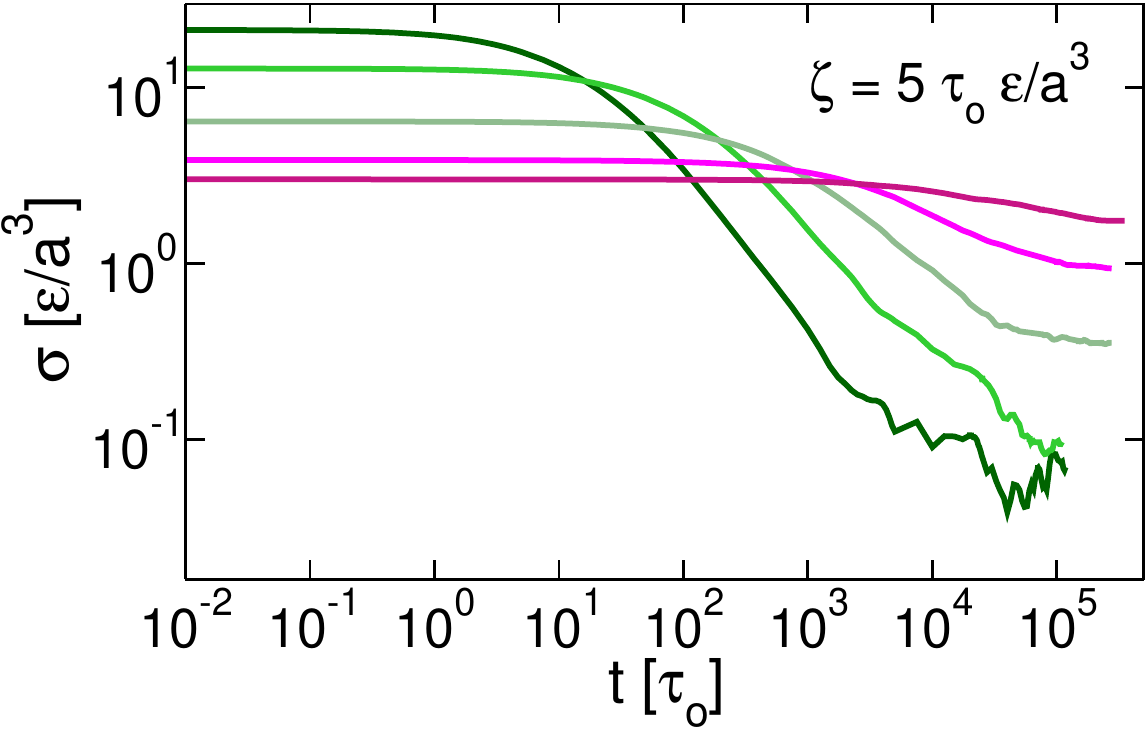}
\includegraphics[scale=0.28]{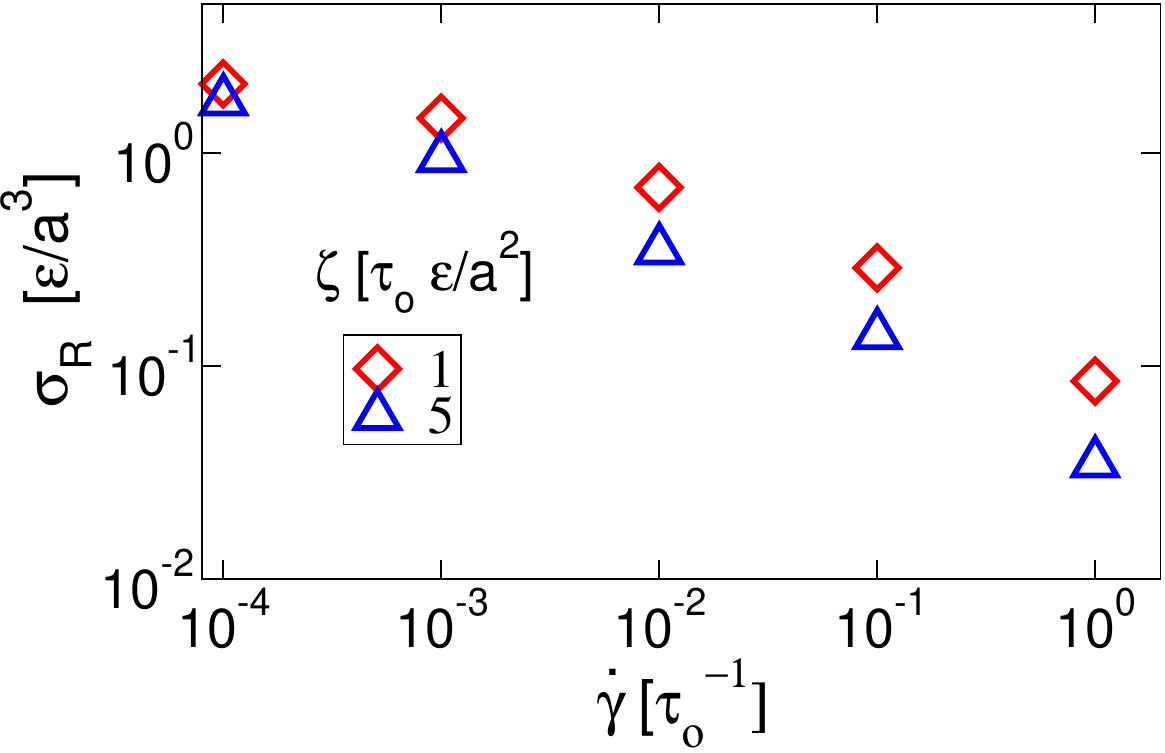}
\caption{\label{stress_relax} Same data as that shown in Fig. 1(b) and (d) in the paper, here graphed without normalization. The data from experiments and simulations are shown in respectively the top and bottom row. The first two rows displays the stress relaxation data obtained for different preshear rates and different solvent viscosities, respectively damping factors. The last row displays the preshear rate dependence of the residual stress for different solvent viscosities, respectively damping factors.}
\end{figure*}

%In Fig. \ref{stress_relax}, we show the evolution of stress and the residual stress for different shear rates and solvent viscosities, upon flow cessation, obtained from both experiments and simulations. The corresponding scaled plots are shown in Fig. 1(b)(c) of the paper. 

\section{Isotropic mean squared particle displacements}

\begin{figure*}[h!]
\centering
\includegraphics[scale=0.5]{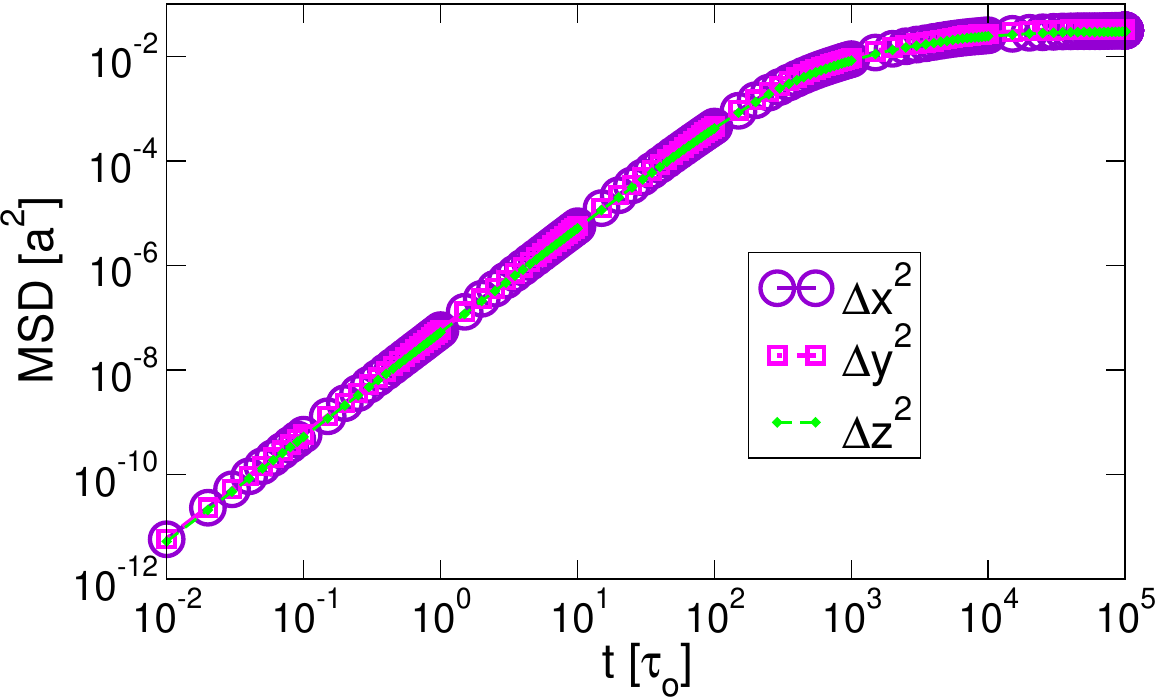}
\caption{\label{msdall} Particle mean squared displacements (MSD) along the flow $X$, the gradient $Y$ and the vorticity $Z$ direction obtained for $\zeta = 1 \uptau_o \epsilon/a^2$ and $\dot{\gamma} = 10^{-2} \tau_o^{-1}$.}
\end{figure*}

As already shown in previous work \cite{mohan} the mean squared particle displacements appears to be isotropic. As an example we show in Fig. \ref{msdall} the mean squared displacement obtained along the flow $X$, the gradient $Y$ and the vorticity $Z$ direction for $\zeta = 1 \uptau_o \epsilon/a^2$ and $\dot{\gamma} = 10^{-2} \tau_o^{-1}$ . The different data sets are essentially indistinguishable. The corresponding total mean squared displacement is shown in Fig. 2(a) in the paper. 

As denoted in our work the mean particle displacement appears to be isotropic because of the coexistence of many domains of highly correlated displacements, whose average displacement vectors point in random directions.\\

\section{Variations in spatial correlations of particle dynamics}
\begin{figure*}[h!]
\includegraphics[scale=0.55]{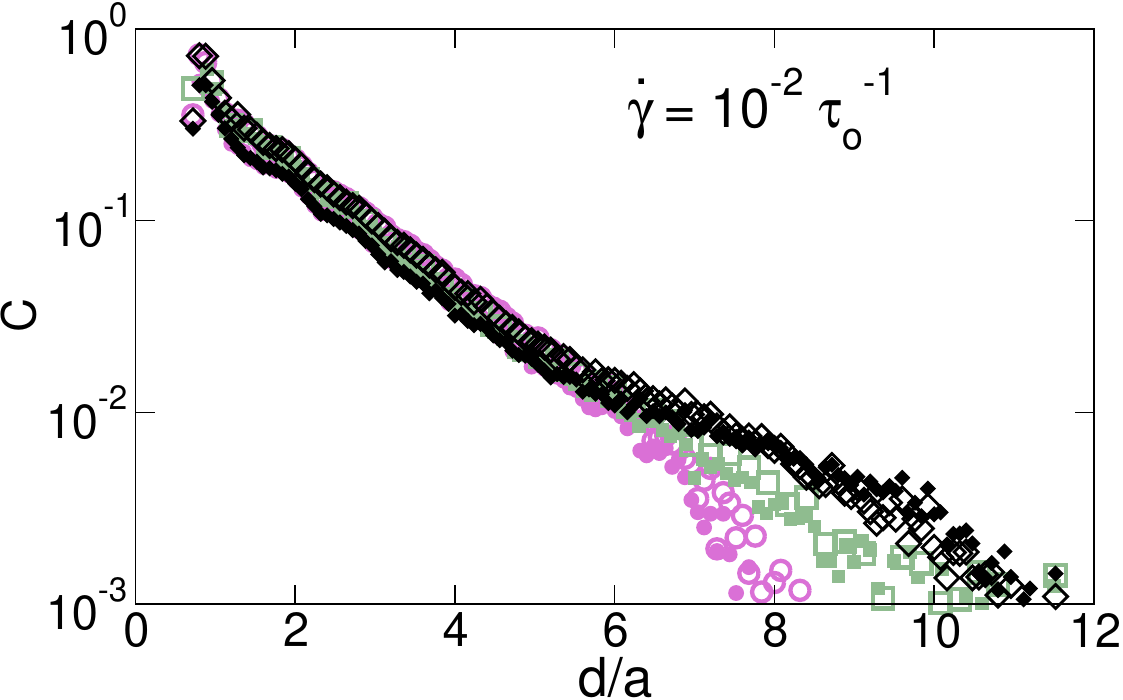}
\caption{\label{corel3} Spatial correlations of particle displacement fluctuations obtained for 3 independent runs with $\zeta = 1 \tau_o \epsilon/a^2$ and $\dot{\gamma} = 10^{-2} \tau_o^{-1}$. Data obtained during flow just before flow cessation are denoted as closed symbols, data obtained upon flow cessation are denoted as open symbols.} 
\end{figure*}
 Performing different runs of a given experiment reveals that the spatial correlations of particle displacement fluctuations $C$ somewhat varies at large distances from one run to another; this is shown for the example of $\zeta = 1 \tau_o \epsilon/a^2$ and $\dot{\gamma} = 10^{-2} \tau_o^{-1}$ in Fig. S3. The agreement of data acquired just before and after flow cessation, however, remains excellent, revealing that the flow cessation characteristics depend on the exact imprint of the preceding flow state. It is also noteworthy that the initial decay used for the determination of the correlation length is nicely reproducible.\\

\section{Structural evolution during stress relaxation}
\begin{figure*}[h!]
%\centering
\includegraphics[scale=0.42]{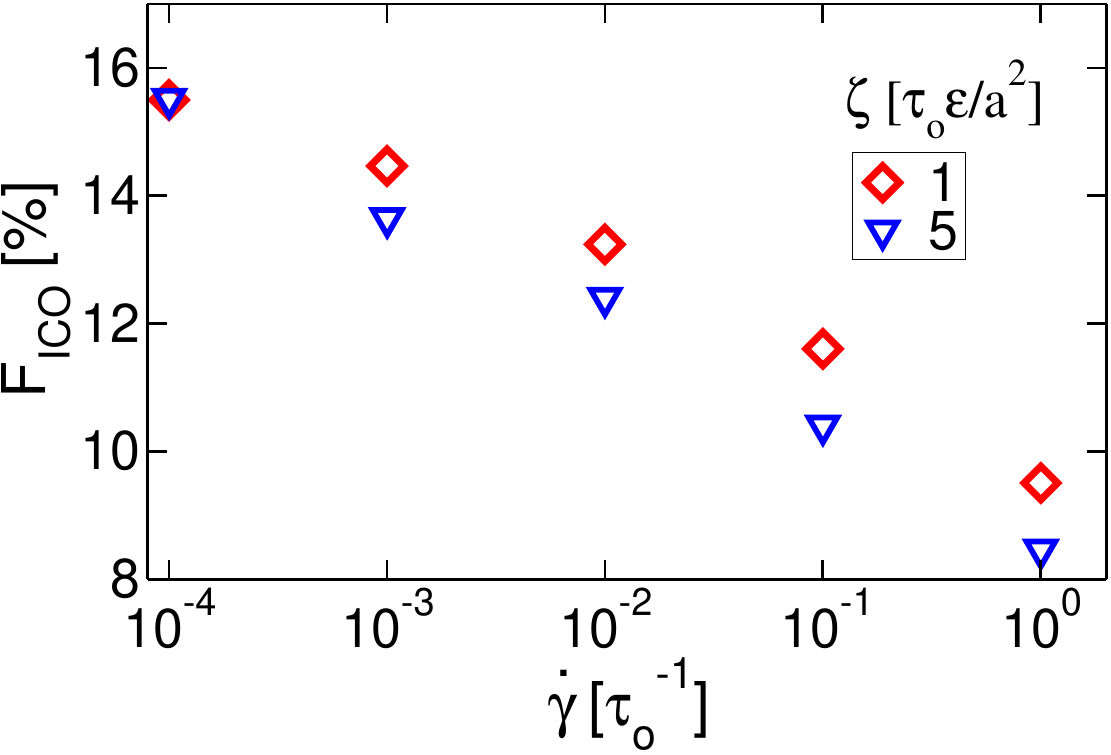}
\includegraphics[scale=0.435]{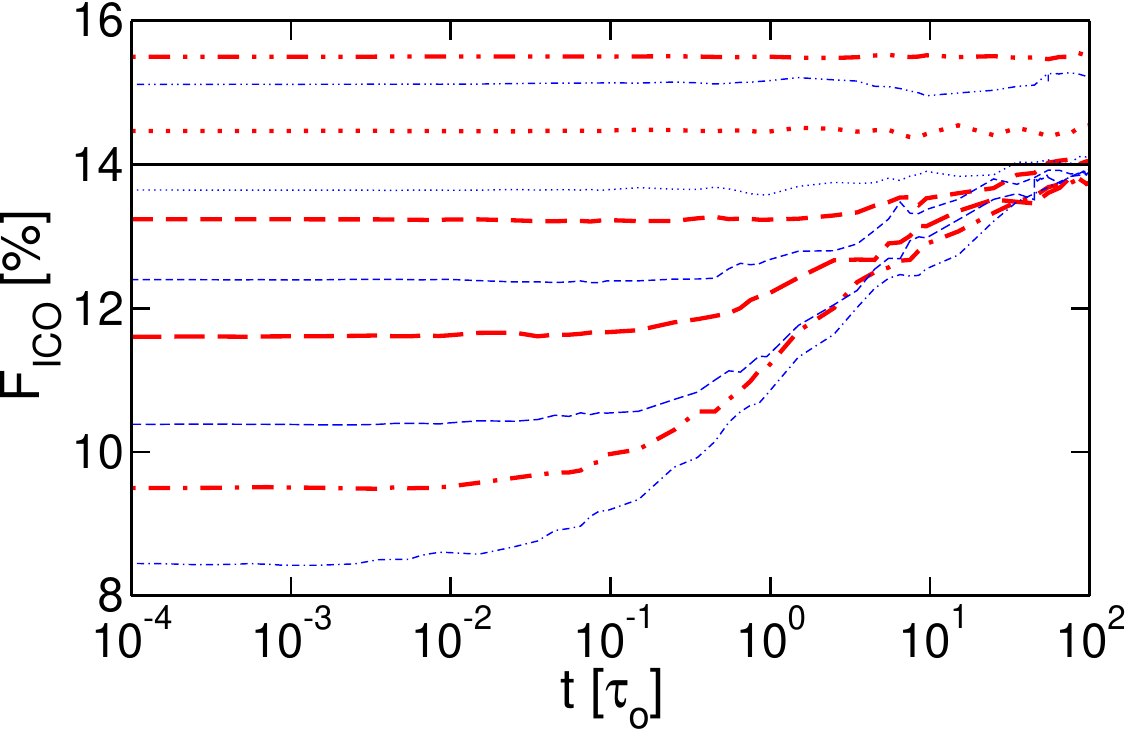}
\caption{\label{delfico} Left panel: Shear rate dependence of the fraction of icosahedral configurations obtained at steady flow conditions. This is the same data as that shown in the main graph of Fig. 4(a), here graphed without normalization. Right panel: Temporal evolution of $F_{ICO}$ upon flow cessation obtained for  $\zeta = 1$ and $5$  $\uptau_o \epsilon/a^2$. From top to bottom: the preshear rate is $\dot{\gamma} = 10^{-4}, 10^{-3}, 10^{-2}, 10^{-1}, 10^{0}$ $\uptau_o^{-1}$. The red and blue lines denote the data obtained for respectively $\zeta = 1$ and $5$  $\uptau_o \epsilon/a^2$.  The black horizontal line marks the magnitude of $F_{ICO}$ reached at the end of the stress relaxation process for the larger preshear conditions, for which we observe $F_{ICO}$ to evolve during stress relaxation.}
\end{figure*}
In the left panel of Fig. \ref{delfico}, we show the fraction of icosahedral configurations ($F_{ICO}$) obtained at steady flow conditions as a function of shear rate instead of the viscous stress, the latter being shown in Fig. 4(a) of the paper. An increase of $F_{ICO}$ during stress relaxation is only observed when the initial value of $F_{ICO} < 14 \%$, as shown for both damping factors investigated in the right panel of Fig. \ref{delfico}. Interestingly, in all these cases $F_{ICO} \approx 14 \%$ is reached at the end of stress relaxation. However, for the condition where the initial $F_{ICO} > 14 \%$ obtained for lower preshear rates, we find that some rearrangements of the icosahedron in space persist upon flow cessation even though $F_{ICO}$ does not increase. To demonstrate this we show the movies of the Voronoi analysis of the structural evolution obtained upon flow cessation after a preshear with respectively $\dot{\gamma} = 10^{-4}$ and $10^{0}$ $\uptau_o^{-1}$ in movie 1 and 2.

\section{Estimation of the volume fraction of the experimental system}\label{secA1}
\begin{figure*}[h!]
\includegraphics[scale=0.6]{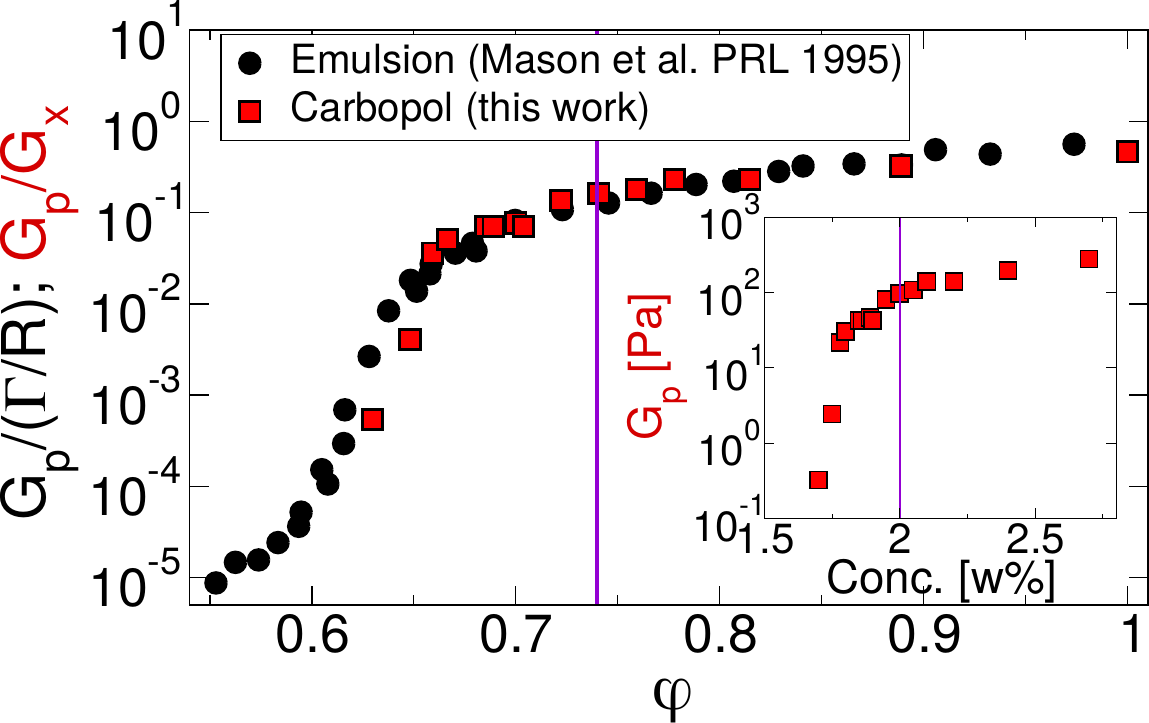}
\caption{\label{figS5} Mapping procedure used to determine the volume fraction of the dispersion of $2 w \%$ Carbopol in propylene glycol. Inset: Concentration dependence of low frequency plateau modulus of Carbopol dispersions in propylene glycol with concentrations indicated in weight percent. The vertical line denotes the experimental system used in this work. Main Figure: Volume fraction dependence of plateau modulus of emulsions (black circles) published in Ref.\cite{mason}. The plateau modulus is here normalized by the ratio of the surface tension and the radius of the emulsion droplets. The red squares denote the data obtained for the Carbopol dispersions (inset) that have been normalized so to match the emulsion data. The vertical line denotes the concentration of the experimental system used in this work.}
\end{figure*}

To estimate the volume fraction $\phi$ of our experimental system, we prepare a series of Carbopol samples with different weight concentrations. For these samples we determine  
the frequency dependence of the storage and loss modulus as a function of frequency in oscillatory strain experiments using a strain amplitude within the linear range. From this data we extract the low frequency modulus $G_p$, which we find to exhibit a concentration dependence reminiscent of that observed by Mason et al. for an emulsion system \cite{mason}.  As shown in the inset of Fig. \ref{figS5}, $G_p$ increases strongly within a narrow range of concentration, to then increase only moderately in the range of larger concentration. The corresponding data set obtained for emulsions as a function of volume fraction is shown as black circles in the main Figure, where $G_p$ is here normalized by the Laplace pressure $\mathit{\Gamma}/R$ with $\mathit{\Gamma}$ the surface tension and $R$ the radius of the emulsion droplets. We estimate the volume fraction of our system by mapping our data to those obtained by Mason et al., normalizing both the modulus and the concentration so to match the emulsion data, as shown in the main Figure of Fig. \ref{figS5}. From this mapping procedure we estimate that the volume fraction of our system with a concentration of $2 w \%$ is about $74\%$  marked by a vertical line in Fig. S5.

%\vspace{10cm}